\documentclass[twocolumn]{aastex63}

\usepackage{amsmath}

\newcommand\g{\; {\rm g}}
\newcommand\pcc{\;{\rm cm}^{-3}}
\newcommand\Msun{\; M_{\odot}}
\newcommand\kms{\; {\rm km}\;{\rm s}^{-1}}

\newcommand\pc{\;{\rm pc}}

\newcommand\surfunit{\Msun\;{\rm pc^{-2}}}

\newcommand\simgt{\lower.5ex\hbox{$\; \buildrel > \over \sim \;$}}
\newcommand\simlt{\lower.5ex\hbox{$\; \buildrel < \over \sim \;$}}

\newcommand\abrackets[1]{\left\langle{#1}\right\rangle}

\newcommand\SFR{\mathrm{SFR}}
\newcommand\mff{{M/t_{\mathrm{ff}}}}
\newcommand\tff{t_{\mathrm{ff}}}
\newcommand\eff{\epsilon_{\mathrm{ff}}}
\newcommand\tffi[1]{t_{\mathrm{ff},#1}}
\newcommand\effi[1]{\epsilon_{\mathrm{ff},#1}}
\newcommand\nh{n_{\mathrm{H}}}

\newcommand\seff{\sigma_{\Delta\SFR/\langle\SFR\rangle}}
\newcommand\tdyn{t_{\mathrm{dyn}}}
\newcommand\Esubd[1]{{\cal E}_{\rm #1}} 
\newcommand\Esubt[1]{E_{\rm #1}} 
\newcommand\nhmin{n_{\rm H,min}}

\newcommand{\comment}[1]{}
\newcommand{\vect}[1]{\mathbf{#1}}
\newcommand{\mean}[1]{\abrackets{#1}}

\newcommand{\figformat}[1]{#1}
\newcommand{\figwell}{\figformat{well}}
\newcommand{\figprogression}{\figformat{progression}}
\newcommand{\figtree}{\figformat{tree411}}
\newcommand{\fighistbin}{\figformat{histbin}}
\newcommand{\fighistthresh}{\figformat{histthresh}}
\newcommand{\fighistx}{\figformat{histx}}
\newcommand{\figav}{\figformat{avc}}
\newcommand{\figpbound}{\figformat{pbound}}
\newcommand{\figlws}{\figformat{lws}}
\newcommand{\figtimemass}{\figformat{timemass}}
\newcommand{\figmbtsfr}{\figformat{mbtsfr}}
\newcommand{\figtimedensity}{\figformat{timedensity}}
\newcommand{\figtimelag}{\figformat{timelag}}
\newcommand{\figtimeothers}{\figformat{timeothers}}
\newcommand{\figeff}{\figformat{effsigma}}
\newcommand{\figerrorhist}{\figformat{errorhist}}
\newcommand{\figcphne}{\figformat{cphne}}

\begin{document}


\title{Cloud Properties and Correlations with Star Formation in Numerical Simulations of the Three-Phase ISM}

\author[0000-0002-2491-8700]{S. Alwin Mao}
\author[0000-0002-0509-9113]{Eve C. Ostriker}
\author[0000-0003-2896-3725]{Chang-Goo Kim}
\affiliation{Department of Astrophysical Sciences, Princeton University, Princeton, NJ 08544, USA}
\shortauthors{Mao, Ostriker, \& Kim}
\shorttitle{Cloud Properties and Star Formation}
\email{alwin@princeton.edu, eco@astro.princeton.edu,cgkim@astro.princeton.edu}

\begin{abstract}
We apply gravity-based and density-based methods
to identify clouds in numerical
simulations of the star-forming, three-phase interstellar medium (ISM),
and compare their
properties and their global correlation with the star formation rate
over time.  The gravity-based method identifies bound objects,
which have masses $M \sim 10^{3}-10^{4}\Msun$ at densities $\nh \sim 100 \pcc$,
and traditional virial parameters $\alpha_{v} \sim 0.5-5$.
For clouds defined by a density threshold $\nhmin$, the average virial parameter
decreases, and the fraction of material that is genuinely bound
increases, at higher $\nhmin$.  Surprisingly, these clouds can be
unbound even when $\alpha_{v} < 2$, and high mass clouds
($10^{4}-10^{6}\Msun$) are generally unbound.  This suggests
that the traditional $\alpha_{v}$ is at best an approximate measure of
boundedness in the ISM.  All clouds have internal
turbulent motions increasing with size as $\sigma \sim 1\kms(R/\pc)^{1/2}$,
similar to observed relations.  Bound
structures comprise a small fraction of the total simulation
mass, with star formation efficiency per free-fall time $\eff \sim0.4$.
For $\nhmin = 10-100\pcc$,
$\eff \sim 0.03 - 0.3$, increasing with density. Temporal
correlation analysis between $\SFR(t)$ and aggregate mass
$M(\nhmin{};t)$ at varying $\nhmin$ shows that time delays to star formation are  
$t_{\mathrm{delay}}\sim\tff(\nhmin)$. Correlation between $\SFR(t)$ and
$M(\nhmin;t)$ systematically tightens at higher $\nhmin$.
Considering moderate-density gas, selecting against high
virial parameter clouds improves correlation with SFR, consistent with
previous work.  Even at high $\nhmin$, the temporal dispersion in
$(\SFR - \eff M/\tff)/\langle\SFR\rangle$ is $\sim 50\%$, due to the large-amplitude
variations and inherent stochasticity of the system.
\end{abstract}


\section{Introduction}\label{sec:intro}

The interstellar medium is hierarchically structured.  The densest
entities are individual prestellar cores, which generally are found
within filaments or clumps in giant molecular clouds (GMCs) \citep{AndrePPVI,DobbsPPVI}; the GMCs may be part of molecular/atomic
complexes, and are typically found within spiral arms, arm
spurs/feathers, or sheared flocculent features \citep{Elmegreen1980,LaVigne2006}. At any given time within a galaxy, a distribution of
GMCs with various properties exists, and each forms stars according to
the distribution of clumps and cores within it. To understand the
intermediate scale between parsecs and kiloparsecs, the properties of
GMCs must be understood, and it is of particular interest to
investigate whether the characteristics of a GMC may be used to
predict its star formation rate.

There is a long history of characterizing ISM structures in
observations. Molecular lines, dust extinction, and dust emission maps
are used to identify regions with high column density or number
density. These density proxies are a convenient and readily available
way to identify structures, and from these measurements distributions
of cloud sizes and masses can be obtained.
In addition to measuring column densities from molecular or dust
emission, line emission is used to trace velocities of gas, and from
this the kinetic energy content of the structures can be estimated.
For example, based on CO surveys, GMCs in the Milky Way have masses $10^{4}-10^{6} \Msun$, radii between $10-50 \pc$, velocity dispersion between $1-7\kms$, and a linewidth-size relationship of $\sigma_{1D} = 0.9 \kms (R/\pc)^{1/2}$
\citep{Solomon1987,Blitz1993,Heyer2015},
and properties of resolved GMCs in nearby
galaxies are similar \citep{Bolatto2008}.

By combining an estimate of the mass, size, and velocity dispersion,
an estimate of the virial parameter $\alpha_v \equiv 2 \Esubt{k}/|\Esubt{g}|$ (for kinetic energy $\Esubt{k}$ and gravitational energy $\Esubt{g}$) can
be obtained \citep[e.g.][]{Roman-Duval2010,HernandezTan2015}.
Virial parameter estimates from observations typically adopt
$|\Esubt{g}|=3 GM^2/(5R)$
for the gravitational energy, as would apply for an isolated,
uniform-density sphere, where the effective radius is empirically
computed from projected area as $R=(A/\pi)^{1/2}$.  Although the case
of ellipsoidal structures has been considered
\citep{Bertoldi1992},
more general effects from nonspherical cloud geometry are not
generally taken into account (even though the filamentary nature of
the ISM makes many clouds quite elongated); non-sphericity tends to reduce
gravitational binding.  Internal stratification is sometimes taken
into account by assuming a power-law density profile, which can
increase the estimated $|\Esubt{g}|$ by up to a factor $\sim 2$ \citep{HernandezTan2015}.

Based on the simplest spherical estimate, clouds are considered
``bound'' if the estimated virial parameter $\alpha_v \equiv
5 \sigma^2 R/(GM)$ is less than or equal to 2, where $\sigma$ is the
line-of-sight velocity dispersion.  However, traditional estimates of gravitational 
binding energy are problematic even beyond the assumptions of
homogeneity and spherical geometry because the ``isolated cloud''
estimate of $|\Esubt{g}|$ does not properly take into account neighboring
structures.  For a given local gravitational potential minimum at the
center of a cloud, tidal forces set the effective zero of the
gravitational potential not at infinite distance but along the first potential
contour that has a saddle point -- equivalent to the Roche
lobe for the case of two spherical bodies. As a result, tidal forces
effectively decrease gravitational binding energy $|\Esubt{g}|$ of dense
regions in close proximity to other dense regions, which is common
because of the hierarchical structure of density variations. In
addition, simple virial parameter estimates neglect magnetic
contributions to support, which can significantly add to the numerator
\citep{McKee1993PPIII,Heiles1993PPIII}.
Although
simple virial parameter estimates are inexact, they are often used to
assess whether a structure is a likely candidate for star formation.

Star formation is observed to take place within the densest structures at the
smallest scale within the ISM hierarchy, and it is important to
understand what dynamical processes lead to the onset of gravitational
collapse, and what controls the rate of star formation within a given
level of the hierarchy.  More generally, it is of interest to
understand how star formation timescales are related to the properties
and corresponding timescales of gaseous structures.  Because star
formation involves gravity, the most commonly invoked reference
timescale is the free-fall collapse time,
\begin{equation}\label{eq:tff}
  \tff = \left(\frac{3\pi}{32G\rho}\right)^{1/2},
\end{equation}
where $\rho$ is the gas density.  Perhaps the simplest way to
characterize the relationship between star formation and gas
properties is via the star formation efficiency per free-fall time
\citep{Krumholz2005,Krumholz2007},
defined as
\begin{equation}\label{eq:eff}
  \eff \equiv \frac{\dot M_*}{M/\tff},
\end{equation}
where $\tff$ is the free-fall time at the mean density of the gas
contributing to $M$, and $\dot M_*$ is the star formation rate (SFR).
Other relevant timescales include the flow crossing time across a
structure that is supported by turbulent stresses, and the sound
crossing time for a structure that is supported by thermal pressure.
A class of theoretical models for star formation suggests that in
turbulent clouds, there is a critical density $\rho_\mathrm{crit}$
above which collapse occurs
within a free-fall time, with $\rho_\mathrm{crit}$ depending on the
the ratios of kinetic to gravitational energy (virial parameter),
turbulent to thermal velocity (Mach number),
and thermal to magnetic pressure (plasma beta parameter)
\citep{Krumholz2005,Padoan2011,Hennebelle2011,Federrath2012,Padoan2014}.
The underlying physical concept behind the idea of a
critical density is that the density must be high enough that thermal
pressure and magnetic stresses cannot support against collapse, and
that the collapse time is shorter than the timescale for shear to
tear apart a structure.

In addition to theoretical models, direct numerical simulations have
been used to characterize the dependence of SFRs on
gas properties.  One idealized type of setup employs simulations with
isothermal, self-gravitating gas, in which turbulence is driven in
Fourier space.  From a large set of driven-turbulence
simulations, \cite{2012ApJ...759L..27P} suggested that $\eff$ depends
primarily on the ratio of flow crossing time to free-fall time as
\begin{equation}
  \eff \propto \exp(-1.6\tff/\tdyn).
\end{equation}
where $t_{\rm dyn} = R/\sigma_{3D} = R/(\sqrt{3}\sigma_{1D})$ is the flow crossing time for system size $2R$ ($=L$, the simulation box size for \cite{2012ApJ...759L..27P}). For a uniform spherical
cloud, the timescale ratio can be related to the virial parameter by 
\begin{equation}\label{eq:tav}
  \left(\frac{\tff}{\tdyn}\right)^{2}
  = \frac{3\pi^{2}}{40}\alpha_{v};
\end{equation}
thus, these simulations suggest a strong suppression of star formation
at high $\alpha_v$.     

Idealized simulations have the advantage of carefully controlled
conditions, but the disadvantages that the turbulence is driven in an
artificially prescribed manner to maintain a fixed overall turbulent
amplitude, and that the processes leading to cloud formation and
destruction are not followed.  In reality, GMCs form due to a
combination of large-scale ISM flows (including turbulence, shear,
and epicyclic motion) and gravity (both stellar gravity and
self-gravity) that lead to collection of material from a large volume,
as mediated by thermal and magnetic pressure, and a change from atomic
to molecular phase as gas cools.  Turbulence on scales less than
the scale height of the warm-cold ISM likely originates
primarily due to the feedback from young stars \citep{Maclow2004,McKee2007,Elmegreen2004}\footnote{Gravitational instabilities in the
  combined gas-stellar system \citep[e.g.][]{1984ApJ...276..114J,1992MNRAS.256..307R,2001MNRAS.323..445R,2007ApJ...660.1232K} can drive horizontal
  motions at very large  scales,
  as seen in numerical simulations 
  \citep[e.g.][and citations within]{2007ApJ...660.1232K,2008ApJ...684..978S,2009MNRAS.392..294A,2011MNRAS.417.1318D,2012MNRAS.421.3488H,2015ApJ...804...18A}, but generally
  these motions do not reach supersonic amplitudes unless they are
  associated with gravitational collapse.  In addition, turbulence at scales
  less than the disk scale height can be driven by spiral shocks
  and by the magnetorotational instability, but numerical simulations show
  that the corresponding amplitudes are relatively modest in cold gas
\citep[e.g.][and citations within]{2004MNRAS.349..270W,2006ApJ...649L..13K,2007MNRAS.374.1115D,2013MNRAS.430.1790B,2010ApJ...720.1454K,2005ApJ...629..849P,2007ApJ...663..183P}.},
whether
inherited from a GMC's formation stage or produced
internally.  
Considering that GMCs live for at most a few turbulent crossing times
or free-fall times \citep{Kawamura2009,Kruijssen2019},
it is not clear that internal GMC conditions can control star formation in a
way that is entirely divorced from their formation and destruction
processes.

In recent years, (magneto)-hydrodynamic simulations have been used to
follow the star-forming multiphase ISM in kpc-size regions at high resolution.
In these simulations,
massive self-gravitating clouds naturally condense out of the diffuse
gas, and within these clouds localized collapse occurs that represents
star cluster formation \citep{Gatto2017,Iffrig2017,KimOstriker2017,Kannan2018,2018A&A...620A..21C}.
By modeling the return
of energy (representing radiative heating and supernova explosions)
from these cluster particles to their surroundings, a self-consistent,
self-regulated state can be reached in which all thermal phases of the
ISM are represented, and a hierarchy of structures is naturally
created.  While the large-scale time-averaged SFR adjusts such that
feedback provides the energy and momentum needed to maintain overall
equilibrium in the ISM as a whole
\citep{Ostriker2010,Ostriker2011,Kim2011,Kim2013},
the collapse to make individual star clusters depends on local
conditions in overdense clouds. Simulations of this kind present an
opportunity to evaluate the role of gravity in binding ISM structures
that are part of a complex environment, and to assess common practices
for estimating gravitational boundedness.  In addition, simulations of
this kind afford a realistic setting to test theoretical ideas
regarding the role of gravitational boundedness in controlling SFRs.

In this paper, we use a large-scale ISM simulation produced in the
TIGRESS framework \citep{KimOstriker2017}
to characterize the properties
of dense structures and their relationship to star formation.  Our
structural decomposition analysis includes methods that are similar to
typical observational practices, in which objects are defined based on
density or column density.  For sets of objects defined by different
density thresholds, we compute statistics of mass, size, and velocity
dispersion, which allows us to compute ``empirical'' virial parameters
and linewidth-size relations.  We compute both traditional virial
parameters (only kinetic energy) and virial parameters including
thermal and magnetic energy.  In addition, we apply another method of
defining structures based on contours of the gravitational potential
(rather than density contours).  In this method we identify bound
objects as regions where the kinetic, thermal, and magnetic energy are
sufficiently low compared to the gravitational energy (computed
relative to a tidally-defined potential contour).  These two analyses
allow us to relate traditional virial parameter  estimates for objects to
measurements of gravitational binding that directly take into account
nonspherical geometry, internal stratification, and tidal forces.  We
shall show that traditional virial parameter estimates can
significantly under- or over-state the true boundedness of ISM
structures.

To study the relationship between gas and star formation, we use
correlations between the temporal history of the SFR and the mass of
gas in different categories of objects, including objects defined by
density thresholds and objects defined by being gravitationally bound.
In this way, we are able to measure how $\eff$ varies as a function of
density and what $\eff$ is for objects that are gravitationally bound
(also allowing for different treatments of surface terms).  We are
also able to measure time delays between the availability of a mass
reservoir and the star formation burst that it produces.  We use
correlation analysis to quantify the relative predictive power of
different star formation models that depend on the traditional virial
parameter, and on our more sophisticated assessment of gravitational
binding.

The plan of this paper is as follows.  In \autoref{sec:methods} we
describe our analysis methods, including how we identify bound objects
(\autoref{sec:mhbr}), the properties we measure for bound objects and
for density-defined objects (\autoref{sec:mprop}), and how we conduct
time-series correlation analyses (\autoref{sec:mtime});
\autoref{sec:analysissum} summarizes our methods and 
\autoref{sec:tigress} describes the 
primary TIGRESS simulation that we analyze.
\autoref{sec:results} presents an overview of structure (\autoref{sec:resultmaps})
and results of our analyses, including
statistics of object 
properties (\autoref{sec:prop}) and time series correlation studies
(\autoref{sec:rtime}), with a summary of trends in the values of
$\eff$ and levels of correlations for various ways of selecting gas
in \autoref{sec:eff}.
In \autoref{sec:conc}
we summarize our results and discuss connections with other
current theory and observations. 

\section{Methods}\label{sec:methods}

In this paper, we analyze the properties of dense and bound gas
structures, and investigate the relationship between the material in
these structures and the star formation rate, as applied to the
fiducial TIGRESS model described in \citet{KimOstriker2017}.  The
methods we develop, described in some detail here, are quite
general and can be applied to other numerical simulation data.  With
some modifications to allow for projected rather than fully
three-dimensional information, our methods can also be applied to
observed data sets.

We begin by describing methods for identifying objects based on
density isosurfaces or on the gravitational potential in comparison to
the kinetic, thermal, and magnetic energy densities (Section
\ref{sec:mhbr}); additional technical details of the algorithm are described
in Appendix \ref{sec:algorithm}.  
We also describe how we quantify object properties
including mass, size, velocity dispersion, and virial parameter
(Section \ref{sec:mprop}).  We then describe our use of time series to
compare the simulated SFR to the history of mass per free-fall time
for different categories of objects (Section \ref{sec:mtime}); this
involves fitting for optimal time delay and efficiency and Bayesian inference
to test models for the dependence on virial parameter.  Finally, in
Section \ref{sec:tigress} we briefly summarize the numerical
implementation and parameters of the TIGRESS model to which we have
applied our analysis.

\subsection{Bound Objects}\label{sec:mhbr}

To motivate our procedure for identifying gravitationally bound structures,
consider what is bound in the Solar system. For example, a pebble on Earth
is bound to the Earth, the Earth-Moon system, the Sun, and the Galaxy,
but is not bound gravitationally to a nearby pebble.  To make this
determination, both the gravitational potential contours and the relative
velocities of the structures involved are needed; a pebble as well as its
neighbors mutually lie within the Earth's gravitational potential as limited
by the Moon's tidal force, and do not have high enough velocities that
they could find themselves on the Moon or escape entirely from the
Solar system.  Thus, we consider the pebble as part of the Earth.

For application to the ISM and star formation, boundedness can also
have a characteristic scale dependence.  Matter on larger scales tends
to have a higher internal kinetic energy ($\sigma^{2} \propto L$) from
the scaling properties of turbulence, but also tends to be
increasingly bound by the gravitational potential ($GM/R \sim
\rho{}L^{2}$).  Depending on the scale dependence of the density,
there may be a hierarchy of boundedness, with bound structures nested within
other bound structures.

\begin{figure}
  \includegraphics[width=\columnwidth]{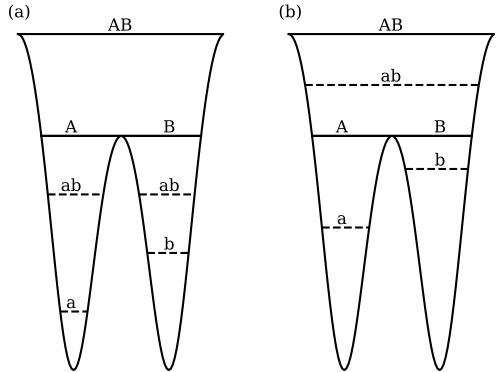}
  \caption{Schematic of HBPs (upper case) and associated HBRs (lower
    case) as level sets within gravitational wells, plotting
    gravitational energy against spatial coordinate. Each HBR is bound
    relative to its associated HBP. For example, on both left and
    right ``a'' represents material interior to an isocontour of the
    gravitational well such that the HBR has 0 total energy, bound
    relative to the gravitational contour ``A.''  Similarly, ``b'' is
    bound relative to ``B.''  The regions ``a'' and ``b'' are
    therefore both HBRs. On the left, we show an example of a region
    ``ab'' which is invalid as an HBR because it consists of two
    non-contiguous parts.  On the right, we show an example of a
    contiguous HBR ``ab'' which merges ``a'' and ``b.''  We are generally
    interested in only the largest HBRs in any hierarchy, so on the right
    we would remove the objects ``a'' and ``b'' from further consideration.
    This schematic also illustrates that each HBP object can be
    identified by a critical point of the potential: ``A'' and ``B'' are
    associated with their respective local minima, and ``AB'' is associated with
    the central local maximum.
  }
  \label{fig:well}
\end{figure}

With the above motivation, we identify a hierarchy of structures in
our ISM simulations based on contours of the gravitational potential,
and bound structures based on the energy of fluid elements relative to
the structure tree.  The first level of the gravitational tree is
comprised of structures enclosed by isocontours that surround a single
minimum.  Each successive level is comprised of material within
isocontours enclosing distinct sets of minima whose largest enclosing
isocontours are in contact. That is, branches merge into a new object
when their isocontours are in contact. This means that each object in
the tree can be uniquely identified with a critical point in the
gravitational potential, where isocontours come into contact.
\autoref{fig:well} provides a schematic illustration of this
procedure.

At each level in the gravitational potential contour tree, we denote the
object enclosed within a closed contour as a {\it hierarchical binding parent}
(HBP).  Within each of these objects, we denote some subset of the gas
as a {\it hierarchical bound region} (HBR).  The HBR is the set of cells for
which the total energy (kinetic, thermal, magnetic, gravitational) of the
region is 0. In this calculation, we assign  a gravitational binding energy
to each cell based on the difference between its gravitational potential ($\Phi$) and the isocontour surface of the HBP ($\Phi_{0}$); i.e. the contribution
to the gravitational energy of the HBR is $(\Phi - \Phi_0)\rho dx^3$. 
Hence, the HBP is responsible for binding these HBR cells.
Cells are added to a candidate HBR in order of gravitational potential depth
(deepest first).  For the contribution to kinetic energy, the center of mass
(COM) velocity of the subset of cells is subtracted out first.  Within a given
HBP, the most massive bound subset of cells is taken as the HBR; if no
cells are bound, there is no HBR.

In the above definition, we have not considered any effects from thermal
or turbulent stresses on the surface (defined by an equipotential) of objects.
Because the dynamics of turbulent systems are complex, surface stresses
could in principle either act to compress and help bind
structures (e.g. for a converging
flow) or act to disperse and unbind structures (e.g. for a shear flow).  
While the complex  dynamics makes it impossible to decide
between these alternatives in a general sense, we can still
investigate the potential magnitude of the effects that
surface stresses may have.
To do this, we begin by averaging the kinetic ($\Esubd{k}$), thermal
($\Esubd{th}$), and magnetic ($\Esubd{B}$) energy density over the
surface $\Omega$ of the HBP ($N$ cells) to compute the mean surface energy density
\begin{equation}
  \Esubd{\Omega} = \frac{1}{N} \sum\limits_{i \in \Omega}
  (\Esubd{k,i} + \Esubd{th,i} + \Esubd{B,i}).
\end{equation}
Here, $\Esubd{k,i}$ is the kinetic energy density computed relative to the
center of mass velocity of the surface cells.

We then define ``HBR+1'' and ``HBR-1'' objects, where
the object HBR$\pm$1 is the set of cells satisfying 
\begin{equation}
  \sum\limits_{i \in \mathrm{HBR\pm1}} \Esubd{k,i} + \Esubd{th,i} + \Esubd{B,i}
   < \sum\limits_{i \in \mathrm{HBR\pm1}} (\Phi_{0} - \Phi_{i})\rho_{i} \pm {\Esubd{\Omega}},
\end{equation}
and now $\Esubd{k,i}$ is the kinetic energy density computed relative to
the center of mass velocity of the HBR$\pm1$ cells.  Clearly, HBR+1
will be more massive than the corresponding HBR identified without the
surface energy terms, because the criterion for including cells
becomes less restrictive by adding $\Esubd{\Omega}$
on the right-hand
side.  Similarly, HBR-1 will be less massive than the corresponding
HBR.  This procedure could be generalized by adding or subtracting
arbitrary multiples of $\Esubd{\Omega}$, but for comparison purposes
we have found HBR-1, HBR, HBR+1 suffice.  We can think of HBR+1
objects as structures in which surface stresses are treated as helping
to bind material; HBR-1 objects are those where surface stresses are
treated as reducing binding.  Physically, addition of $\Esubd{\Omega}$
on the right-hand side in HBR+1 is equivalent to only
considering the excess of $\Esubd{k}$, $\Esubd{th}$, and $\Esubd{B}$
  over ``ambient'' values when computing total
energy.\footnote{We note that when using the Virial Theorem
  \citep[e.g.][]{1992ApJ...399..551M}, in the case of isotropic
  magnetic fields and a spherical cloud the surface terms would enter in 
  exactly the same way as in the HBR+1 definition.  That is, the mean surface
  values of kinetic, thermal, and magnetic energy density would be 
  subtracted from the mean values within the volume.}
 
Subsequently, we will test the correlation of HBR and HBR$\pm$1 objects
with respect to the star formation rate; if surface terms play an important
physical role, we might expect this to be reflected in the relative
correlations with SFRs that we measure.

In the hierarchical contour tree, a nested sequence of HBPs
is uniquely defined by critical points in the equipotential, e.g.
in Figure~\ref{fig:well} ``A'' and ``B'' are nested within ``AB''.
At each level of the tree, HBRs can be identified with respect to the
corresponding HBPs.  
An additional requirement for HBR (and HBR$\pm1$) objects to
be considered valid is spatial compactness. Physically, we impose
this requirement because a ``divided'' HBR within a single HBP could
not be trusted to form a contiguous object.

The case of a non-contiguous HBR occurs when the COM velocity of the
HBP is significantly different from that of its HBP branches, while
the surface potential of the HBP is not significantly higher than that
of its branches (this difference is equivalent to the difference
between the HBP surface and the HBP originating critical point). Then,
in this scenario, considering the HBP as a whole increases the kinetic
energy without sufficiently increasing the depth of the binding
gravitational well, resulting in separate regions that are unbound
relative to each other but may be individually
self-bound. \autoref{fig:well} shows an example of a contiguous
vs. non-contiguous HBR.

Both HBPs and HBRs are grown contiguously from cells of decreasing
potential depth, so a non-contiguous HBR would only form from an HBP
with multiple local minima and correspondingly multiple HBP
branches. The separate HBR subregions are each a subset of cells from
one of those separate HBP branches, which meet at the critical point
identified with the origin of the HBP. Hence, in our tree construction it
suffices to check that the HBR of an HBP contains the critical point
identifying that HBP, which is a convenient guarantee that the HBR is
contiguous.

We have so far described a process of building a contour tree of
gravitational potential isocontours. Each isocontour defines an HBP hosting
an HBR (which may be empty, or may be non-contiguous).
Because HBPs and HBRs are nested, one can consider levels in the hierarchy
separately (in which case given fluid elements are counted at each level
they appear), or one may apply a merging or pruning criterion to objects
to ``flatten'' the hierarchy, such that each fluid element appears in at
most a single object.  

We are interested in regions self-bound on each scale. An HBP may bind a contiguous region of mass but have children binding non-contiguous regions.
This is analogous to a larger scale self-bound system (like a galaxy) containing subregions that are not self-bound (an unbound GMC, with separate bound subregions, would be a non-contiguous HBR). To enforce that every level of the hierarchy is self-bound, we build the HBR tree from the HBP tree bottom-up, starting with leaf HBPs around individual local minima. A parent HBP is only evaluated (computing its HBR) if all its children HBPs were evaluated and host contiguous HBR. If an HBR is evaluated and contiguous, it replaces its branch HBRs, thus becoming a leaf node of the subset of the full HBR tree. The leaf nodes of the subset of the full HBR tree are considered to be star-forming regions. 

This method naturally selects the largest scale candidates for contiguous
collapse and hence star formation, 
and are robust to small-scale fluctuations in
the gravitational potential (for example, the point-potentials of star
particles). Furthermore, we note that changes over time in the
gravitational potential structure can cause rapid changes in the
population of leaf nodes of the contour tree. As objects disperse,
merge, or evolve, critical points are created, destroyed, and
relocated. Leaf nodes can be sensitive to such changes in the
gravitational structure,
but contiguous HBRs
are more robust. For example,
a dispersing or merging object smoothly transitions to or from being
considered as multiple HBRs vs. a single HBR, because the relevant
parameter is the total energy content, which (roughly) continuously
changes.

We also checked the approach of building the HBR top-down, allowing a given HBR to contain non-contiguous HBRs, essentially allowing gas to be bound under any viable isocontour. This method produces contiguous HBRs that are as massive as possible.

Even with the above definitions, additional
choices can be made in computing contiguous HBRs.
We varied a single choice at a time to study their effects.
For example, the star particle potential contribution to the potential
could either be included or excluded. 
However, we found that including the star particle potential did not
produce a significant quantitative effect on our results.
The most significant difference we found was when building the HBR tree
top-down: we found that in certain simulation snapshots (a
few per cent of the time) where the mass
happened to coalesce in a single region, the fraction of ``bound mass''
spiked by an order of magnitude.
For the rest of this paper, when we refer to ``HBR'' the choices adopted
are: building the HBR tree from the bottom-up, excluding the star particle potential, and ignoring surface
stresses.
We have found that that considering surface
stresses can have a large effect,  and we report results separately for
objects identified as HBR$\pm1$, as above.  Inclusion of surface stresses
as HBR+1 can lead to an order of
magnitude more mass being considered ``bound.'' However, as we shall show,
this does not have a
strong effect on the correlation between star formation and ``bound'' mass
over time.

\subsection{Object Definition and Properties}\label{sec:mprop}

After computing HBRs from the gravitational potential structure and
fluid properties (density, kinetic energy, thermal energy, magnetic
energy), we have a collection of objects, each of which contains a
number of simulation cells.

There are also other means of identifying potentially star-forming
objects that are closer to traditional observational methods that use
molecular tracers that may have a characteristic threshold density, or
dust emission/extinction maps with a minimum column.  Here, we shall
apply number density thresholds ($\nhmin=$ 10, 30, and 100 $\pcc$) to
identify contiguous regions where the number density $\nh > \nhmin$, referring to these regions as ``$\nhmin$ objects.''\footnote{For this, we use Python package scipy, specifically the function scipy.ndimage.label, with a boundary correction for the shearing periodic box}
We also treat the set of HBPs for HBRs as objects.

We shall analyze HBRs, HBPs, and $\nhmin$ objects 
in similar ways, both in terms of their properties and
their relation to star formation.  For each set of object categories
in any simulation snapshot, we use member cells to calculate each
object's mass, volume, free-fall time from its mean (volume-weighted)
density, and a mass per free-fall time.

For HBR and $\nhmin$ objects we compute individual virial
parameters $\alpha_{v}$. We compute the thermal energy density
$\Esubd{th} = {\cal P}/(\gamma-1)$ for pressure ${\cal P}$ using
$\gamma = 5/3$.
With momentum density $\vect{p}=\rho \vect{v}$
and center-of momentum velocity
${\bf v}_{\rm COM}$, the kinetic energy density is $\Esubd{k} =(1/2)(
|\vect{p}|^{2}/\rho{} - \rho{}\vect{v}_{\rm COM}^{2})$, while magnetic
energy density is $\Esubd{B} = |\vect{B}|^{2}/8\pi$ for
magnetic field $\vect{B}$.  These are summed over cells for each object to
define the total kinetic, thermal, and magnetic energy $\Esubt{k}$,
$\Esubt{th}$, and $\Esubt{B}$, respectively. 
We define an effective object radius $R$ from
each object's volume via $V = (4\pi{}/3)R^{3}$, and then define an
estimated gravitational self-binding energy as
\begin{equation}\label{eq:Egdef}
  \Esubt{g} \equiv
  \frac{(3/5)GM^{2}}{R}
\end{equation}
using the total object mass $M$.  We note that this is
{\it not} the true gravitational binding energy, but we adopt this
definition for the purpose of comparison with standard practices in
the field which assume isolated objects.  With the above definitions, we set
\begin{align}\label{eq:virialdef}
  \alpha_{v} &\equiv 2\frac{\Esubt{k}}{\Esubt{g}} \\
\label{eq:virialtotdef}
\alpha_{\rm v,total} &\equiv 2\frac{\Esubt{k} + \Esubt{th} + \Esubt{B}}{\Esubt{g}};
\end{align}
  while the former
is used most often in the literature under the assumption that kinetic energy
dominates over both thermal and magnetic, the latter is more general.  
We also examine the separate energy components of objects.  

For each $\nhmin$ object, we find the mass fraction of
cells that are also within HBRs. This allows us to examine the overlap
between a method of identifying ISM structures (and possible
star-forming regions) that is simple but easily applied, and a method
that is sophisticated and physically motivated, but not directly
applicable in observations. This mass
fraction is also the probability of gas being bound given the observation
that it is high density ($P(\rm bound | \rm dense)$). We colloquially
refer to this as the ``bound fraction.''

\subsection{Time Series Analysis}\label{sec:mtime}

A question of significant interest is the detailed correlation in time
between the mass in identifiable star-forming structures and the
actual star formation rate (SFR).  To investigate this question, for each
simulation snapshot and object type defined as described in
Sections~\ref{sec:mhbr} and \ref{sec:mprop}, we sum the mass, volume,
and mass per free-fall time of all objects of that type within the
snapshot. This procedure provides a set of time series (representing the
ratio of mass per free-fall time for selected gas subsets) that we can
use to test the connection to time-dependent star formation rate
$\SFR(t)$.

To create time series for comparison to $\SFR(t)$, we also consider the
collective material above minimum gas surface density thresholds
$\Sigma_0=$ 10, 30, and 100 $\surfunit$.  For each threshold and for
each simulation snapshot, we compute the mass above the threshold,
the volume ($\nhmin$ objects only), and
mass per free-fall time. For the free-fall time, we use a
mass-weighted average density.  For logarithmic bins of number density
of half-decade width, we also compute the snapshot mass, volume, and
mass per free-fall time, using the volume-weighted average
density. This average density tends to be the lower edge of the bin
when looking at the high density side of the distribution.

We compute the SFR at any given time $t$ by taking the total mass of
all star particles whose age $t_{*}$ is less than some maximum age
$t_{*,\rm max}$, and dividing by that age:
\begin{equation}\label{eq:sfrage}
 \rm{SFR}(t) = \sum\limits_{t_{*} < t_{*,\rm max}}M_{*}/t_{*,\rm max}
\end{equation}
This is observationally motivated but also naturally smooths the SFR
time series. This also introduces a delay shift of $t_{*,\rm max}/2$
in the time series relative to the simulation because mass that has
formed stars at a given time $t$ contributes equally to the SFR at
later times until $t+t_{*,\rm max}$, with midpoint centered on
$t+t_{*,\rm max}/2$. As long as only young stars are considered and
$t_{*,\rm max}$ is small, these effects are not problematic.

We use time series comparisons to
compute the star formation efficiency per free-fall time
(\autoref{eq:eff}) for each subset of the gas. For comparison to $\mathrm{SFR}(t)$,
we use the individual time series $\mff$ from each  defined gas
subset (e.g. HBR, HBP, number density thresholds,
number density bins, surface density threshold). Note that the typical
density and free-fall time of a given definition do not significantly
change over time, so correlating $\mathrm{SFR}$ with total mass per free-fall
time $\mff$ is similar to correlating $\mathrm{SFR}(t)$ with the total mass
$M(t)$ in a defined subset.

We treat our time series of simulation snapshots as a set of 2-D
samples in $\mathrm{SFR}$ and $\mff$, and apply a simple linear regression to
estimate $\eff$ using the model $\mathrm{SFR} = \eff\mff$. Hence, our
inferred $\eff$ is simply
\begin{equation}\label{eq:effinfer}
\eff = \frac{\sum_{N} \SFR_{i} (\mff)_{i}}{\sum_{N} (\mff)_{i}^{2}} = \frac{\langle(\SFR)\mff\rangle}{\langle (\mff)^{2} \rangle}
\end{equation}
with standard error in the fitted coefficient
\begin{equation}\label{eq:effstderr}
\Delta\eff{}^{2} = \frac{1}{N-2}\frac{\sum_{N} \Delta{}\SFR_{i}^{2}}{\sum_{N} \left(\mff_{i} - \langle\mff{}\rangle\right)^{2}}
\end{equation}
where
\begin{equation}\label{eq:deltaSFR}
  \Delta\SFR_{i} = \SFR_{i} - \eff{}\mff_{i}
\end{equation}
  and
$\langle\mff\rangle = (1/N)\sum_{N}(\mff)_{i}$.
We note that the uncertainty in  the inferred $\eff$  from
\autoref{eq:effstderr} will tend to decrease with increasing sample size.

The normalized variance in the error of the ``data'' SFR compared to the
``model'' $\eff M/\tff$ is then
\begin{equation}\label{eq:SFRerr}
  \sigma_{\Delta \SFR/\langle \SFR \rangle}^{2} = \frac{1}{N-1}\frac{\sum_{N}
    \Delta\SFR_{i}^{2}}{\langle  \SFR \rangle^{2}}.
\end{equation}
We interpret a smaller normalized error
$\sigma_{\Delta \SFR/\langle \SFR \rangle}$
as a stronger relationship between SFR and
$\mff$. Note that the covariance and Pearson correlation coefficient
between two variables $X$ and $Y$ both increase with a term
$E[XY]$. The standard error increases with
$\Delta\SFR_{i}^{2} = \SFR_{i}^{2} + \eff^{2}(\mff)_{i}^{2} - 2\eff\SFR_{i}(\mff)_{i}$, hence
decreasing with $E[(\SFR)\mff]$. Thus, qualitatively, a smaller standard
error corresponds to a larger covariance or correlation coefficient,
and demonstrates a stronger dependence of SFR on $\mff$.

The above approach has the strength of being a statistically
uncontroversial way to consistently estimate both $\eff$ and the
strength of the relationship between SFR and $\mff$. It also benefits
from giving more weight to simulation snapshots with more $\mff$ or
equivalently more total mass. This is similar to treating each unit of
mass as a single sample and averaging these
samples. Conceptually, it is best to measure $\eff$ in snapshots where
both SFR and $\mff$ are large.

We note that we also experimented with other methods of
estimating $\eff$ and quantifying the connection between SFR and
$\mff$, but use \autoref{eq:effinfer} and \autoref{eq:SFRerr} because these methods
have the most statistical simplicity, physical motivation, and
consistent results of our candidate methods. Other ways of estimating
$\eff{}$ included $\mean{\SFR}/\mean{\mff}$ and
$\mean{\SFR/(\mff)}$. Other ways of quantifying the connection included
the covariance, the Pearson correlation coefficient, the standard
deviation of $[\SFR/(\mff)]_{i}$, and the root mean square of $\SFR -
\eff{}\mff$.

In practice, we modify the above to consider the effect of time
delays.  First, as already alluded to, our definition and observable
definitions of SFR are already shifted. Furthermore, given a gaseous
object it is reasonable to expect that it may not presently be forming
stars but rather will form stars after a delay which scales with the
free-fall time.  In detail, we might expect that temporal peaks in the
mass of low density gas would lead to temporal peaks in the mass of high
density gas after a delay comparable to the low density free fall
time.  Correspondingly, temporal peaks in the mass of gas at yet
higher density might be expected after a subsequent shorter delay,
comparable to the high density free fall time.

To allow for temporal delays, we apply the analysis described by
\autoref{eq:effinfer} and \autoref{eq:SFRerr} to time-shifted sets of SFR and
$\mff$, interpolating when necessary.
For any time series, we identify the delay time $t_{d}$ which
minimizes
$\sigma_{\Delta \SFR/\langle \SFR \rangle}$,
assuming that SFR lags behind $\mff$
by $t_{d}$.  We present results for $\eff{}$ and
$\sigma_{\Delta \SFR/\langle \SFR \rangle}$
for this choice of $t_{d}$.  This allows for the maximum correlation
between SFR and $\mff$, under the assumption that $\mff$
causes future SFR.


\subsubsection{Dependence on virial parameter}\label{sec:effmodels}

In varying galactic environments, gas at a given density may be in
different dynamical states, in ways that would affect future star
formation.  For example, increasing contrast of the density in a cloud
relative to its environment may reflect a more bound state, and clouds
that are more bound might be more susceptible to forming stars.  
Following typical practice in the field, we can characterize the ``boundedness''
of individual structures based on their virial parameter.

We test
the effect of the virial parameter on susceptibility to star formation
using our time series, comparing the actual $\SFR(t)$ (from star particles)
with model predictions
\begin{equation}\label{eq:modelsfr}
  \SFR_{m}(t) = \sum\limits_{\mathrm{object\  i}} \frac{\eff(\alpha_{v,i}) M_{i}}{\tffi{i}}.  
\end{equation}
For each temporal snapshot, the right-hand side is a sum over objects in
a given category at that time (with ``objects'' being HBRs or density-defined objects), and
$\eff(\alpha)$ is a specified model.  For each object, 
$\alpha_{\rm v,i}$, $M_i$, and $t_{\rm ff,i}$ are the virial parameter, mass, and
free-fall time.  

Our simplest model is to take constant $\eff$, that is
\begin{equation}
  \eff(\alpha_{v}) = \effi{0}.
\end{equation}
where $\effi{0}$ defines the normalization of this model and models to follow.

Our second model is a generalization of the dependence proposed by 
\citet{2012ApJ...759L..27P},
\begin{equation}\label{eq:padoanmodel}
  \eff(\alpha_{v}) = \effi{0} \exp(-\beta(3\pi^{2}/40)^{1/2}\alpha_{v}^{1/2}).
\end{equation}
In their simulations of self-gravitating driven-turbulence periodic
boxes for models with a range of global ratios of $\tff/t_{\rm dyn}$,
they found that $  \eff =\effi{0} \exp(-\beta \tff/t_{\rm dyn})$ with
$\beta = 1.6$ followed the overall trend for the dependence of
SFR on the value of $\tff/t_{\rm dyn}$ (see Section\ref{sec:intro}).

Another possible model is a simple $\alpha_{v}$ cutoff,
\begin{equation}\label{eq:cutoffmodel}
  \eff(\alpha_{v}) = \effi{0} H(\alpha_{v,\mathrm{cutoff}} - \alpha_{v}),
\end{equation}
where H is the Heaviside step function, thus taking only objects with $\alpha_{v} < \alpha_{v,\mathrm{cutoff}}$ but weighting them equally. 

Since we are interested in comparing model $\SFR_{m}(t)$ to simulation
$\SFR(t)$, we apply Bayes's theorem,
\begin{equation}\label{eq:Bayes}
  P(A|B) = \frac{P(B|A)P(A)}{P(B)}
\end{equation}
where A represents the model $\SFR_{m}(t)$ given by \autoref{eq:modelsfr} and B represents the simulated SFR(t) given by \autoref{eq:sfrage}. 

For the likelihood $P(B|A)$ we assume
\begin{equation}\label{eq:likely}
  P(B|A) = \prod_{i} \frac{1}{\sqrt{2\pi\sigma^{2}}} e^{-\frac{(\Delta\SFR(t_{i})/\langle\SFR\rangle)^{2}}{2\sigma^{2}}},
\end{equation}
taking the product over discrete time samples $t_{i}$, and where
\begin{equation}
  \Delta\SFR(t) = \SFR(t) - \SFR_{m}(t-t_{d}).
\end{equation}
We select subsets of $\{t_{i}\}$ for each delay time $t_{d}$ so that the
likelihood $P(B|A)$ is always computed using the same number of
samples/snapshots regardless of $t_{d}$. We normalize by the time
averaged global star formation rate $\langle\SFR\rangle$ so that
$\sigma$ is dimensionless.

For a given object class and model $\eff(\alpha_{v})$, we evaluate the
likelihood $P(B|A)$ over the parameter vector $\theta$ that includes
time delay $t_d$, $\effi{0}$,  additional model parameters ($\beta$ or
$\alpha_{v,\mathrm{cutoff}}$ as appropriate), and $\sigma$.
Since $A$ represents SFR$_m$ and depends only on the parameter vector
$\theta$, the posterior in \autoref{eq:Bayes} is
\begin{equation}
  P(\theta|\SFR) = \frac{P(\SFR|\theta)P(\theta)}{P(\SFR)}.
\end{equation}
Note that $P(\theta) = \prod_{i} P(\theta_{i})$, the product of
priors, which we briefly describe.  We use uniform linear priors for
time delay $t_{d}$ and slope $\beta$ (allowing negative values), and
uniform logarithmic priors for $\effi{0}$, $\sigma$, and 
$\alpha_{v,\mathrm{cutoff}}$. Using uniform linear priors instead of logarithmic does
not substantially change our results.

Since $P(\SFR)$ does not vary with $\theta$, we estimate the
marginalized distribution for parameter $x$ by integrating over other
parameters $\Theta = \{y \in \theta | y \neq x\}$
\begin{equation}
  P(x|\SFR) = \frac{\int P(\theta|\SFR) d\Theta}{\int P(\theta|\SFR) d\theta} = \frac{\int P(\SFR|\theta) P(\theta) d\Theta}{\int P(\SFR|\theta) P(\theta)d\theta}
\end{equation}
thus inferring mean values of each parameter
\begin{equation}\label{eq:Bayesmean}
  \hat{x} = \int x P(x|\SFR) dx
\end{equation}
and variance from
\begin{equation}\label{eq:Bayesvariance}
  \mathrm{Var}(x) = \hat{x^{2}} - \hat{x}^{2}
\end{equation}

For the constant $\eff$ model, inferring $\eff$ is equivalent to obtaining
the simple linear regression described above in Equation~\ref{eq:effinfer}.

From the definition of $\sigma$ in \autoref{eq:likely}, the inferred
value of $\sigma$ is equivalent to $\sigma_{\Delta \SFR/\langle \SFR
  \rangle}$, and is a measure of the goodness of fit of each model to
the data for the inferred parameter values.  Beyond the single value
of $\sigma$, it is also interesting to compare the distributions in
$\Delta \SFR_i/\langle \SFR\rangle$ for different models and different
gas subsets.

\subsection{Analysis Methods Summary}\label{sec:analysissum}

To summarize, we compute two categories of properties from a set of
simulation snapshots: object-by-object properties of all objects from
all snapshots, and time series of simulation snapshot totals.
Object-by-object properties include mass, volume, mean density,
free-fall time, mass per free-fall time, and virial parameters, and
only apply to number density threshold, HBR, and HBP objects. The
bound fraction is an object-by-object property only applying to $\nhmin$
objects.  Simulation snapshot total properties
include mass, volume, and mass per free-fall time, which we calculate
for all definitions: surface density threshold, number density
threshold, number density bins, HBRs, and HBPs.  We use the
object-by-object properties to study the population of star-forming
objects.  We use the time series to study temporal
correlations between the star formation rate and various subsets of
the simulation gas.

\subsection{TIGRESS Simulations}\label{sec:tigress}

Although our methods can be applied more broadly, we focus our tests
on the TIGRESS simulations. TIGRESS is a suite of MHD simulations
which self-consistently models star formation and effects of feedback
in the three-phase ISM at parsec scales.  Details of the TIGRESS numerical
algorithms are presented in \citet{KimOstriker2017}, along with
results on the basic properties (and a convergence study) of
a model with parameters representative of the Solar neighborhood.  We
use two versions of this  model for the tests in the present paper, one
with 4 pc resolution and one with 2 pc resolution.  Data dumps from these
models that we use have a cadence of 1 Myr, with different minimum and
maximum times as indicated in Table \ref{tab:simparam}.   While the
surface density declines over time, the typical value is $\sim 10 \surfunit$.

The features in TIGRESS include self-gravity, sink particles,
supernova rates and FUV luminosity from a population synthesis model,
resolved supernova remnant evolution prior to cooling,
FUV-dependent photoelectric heating, optically thin cooling, and galactic shear.
TIGRESS uses shearing periodic boundaries in the galactic plane and
outflow in the vertical direction. The shearing periodic boundaries
affect the computation of gravitational potential isocontours. We use
an algorithm wherein each cell only needs to know which cells are its
immediate neighbors, so to correct for shearing periodic boundaries
(or any other boundary) we simply correct the neighbor list of cells
on the boundary. The shear velocity is included in computing the
kinetic energy of objects, but is a small effect. However, a
correction is necessary across a shearing periodic boundary, since an
object lying across the boundary contains cells with extra velocity
$qL\Omega = 28.7 \kms$.  

\begin{deluxetable*}{l l l l l l l}\label{tab:simparam}
  \tablecaption{Simulation Parameters \label{tab:table1}}
  \tablehead{\colhead{Name} & \colhead{Resolution} & \colhead{Cadence} & \colhead{$t_{\mathrm{min}}$ (Myr)} & \colhead{$t_{\mathrm{max}}$ (Myr)} & \colhead{$\Sigma_{\mathrm{min}} (M_{\odot}/\pc^{2})$} & \colhead{$\Sigma_{\mathrm{max}} (M_{\odot}/\pc^{2})$}}
  \startdata
  MHD\_4pc   & 4 pc      & 1                & 300       & 700       & 8 & 13\\
  MHD\_2pc   & 2 pc      & 1                & 351       & 421       & 9 & 10\\
  \enddata
\end{deluxetable*}


\subsection{Dendrograms}

We can use a dendrogram\footnote{See \cite{2008ApJ...679.1338R,2009Natur.457...63G,2013ApJ...770..141B} for previous dendrogram analyses} as a graphical representation of the
gravitational potential contour
tree. Simultaneously, a dendrogram represents the structure of the
gravitational potential and shows where HBRs are relative to that
structure. We compute a dendrogram so that local minima in the
gravitational potential are spaced evenly (``Tree Index'') and ordered
so that two objects that intersect are nearby.
Then, the intersections can be represented by
non-overlapping horizontal lines, and distances in ``Tree Index''
roughly encode 3-D spatial distances, since intersecting isocontours
are obviously in contact with each other. We start with a list of all
isocontours on top of the tree with no parents. A complete contour
tree should only have one such isocontour containing all points, but
it may be desired to terminate the evaluation of the contour tree
early. Then, each member of the list is replaced with itself followed
by its immediate children, the isocontours that merged to form
it. This repeats for each new member of the list and can be performed
recursively. Forming this list top-down keeps children together, with
only their descendents between them, which ensures that intersections
do not overlap. Then, the tree is plotted in reverse, since deeper
descendents appear later in the list and need to be plotted first, as
the average of their ``Tree Index'' determines the ``Tree Index'' of
their parents. Local minima are plotted first and given integer ``Tree
Index,'' which evenly spaces them, as desired.

\begin{figure}
  \includegraphics[height=0.95\textheight,width=\columnwidth,keepaspectratio]{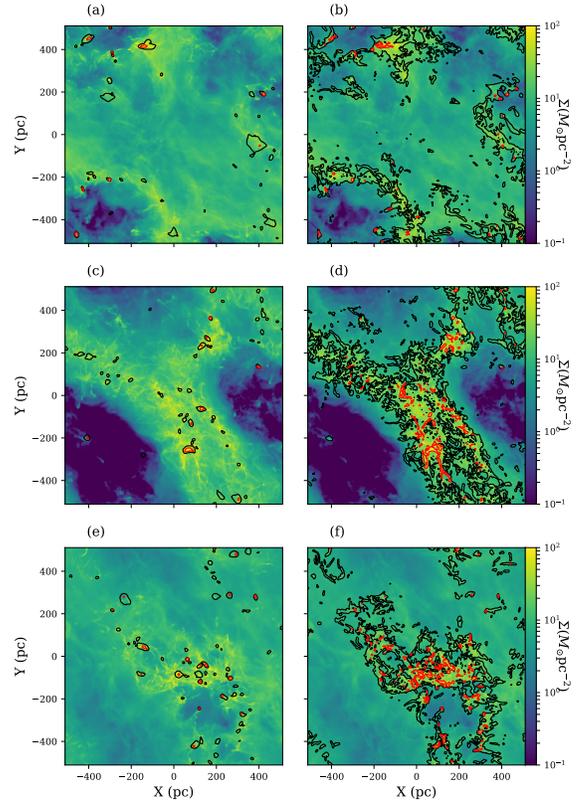}
  \caption{A progression of surface density snapshots from the TIGRESS
    2pc (MHD\_2pc) simulation at 370, 390, and 410 Myr (from top to bottom)
    comparing
    energy-identified objects (left) to density-identified
    objects (right). Red contours show projections of
    HBR (left) and $\nhmin = 100\pcc$ objects
    (right). Black contours show HBP (left) and $\nh > 10\pcc$ objects
    (right).}
  \label{fig:progression}
\end{figure}

\begin{figure}
  \includegraphics[width=\columnwidth]{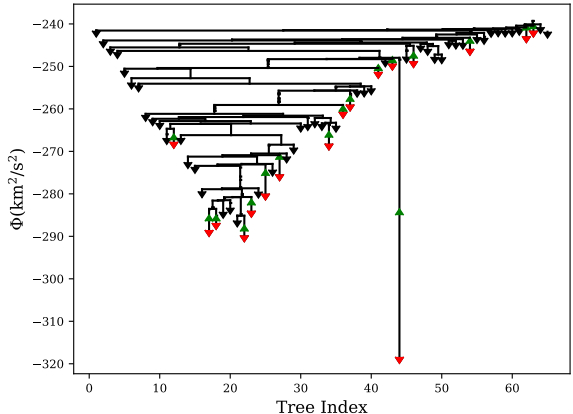}
  \caption{A representation of the contour tree, or dendrogram,
    showing objects according to their gravitational potential value
    ($\Phi$, in units of $(\kms)^{2}$) and relative position in the
    tree. This corresponds to the bottom panels of
    \autoref{fig:progression}, from the 2pc resolution simulation at
    410 Myr. Downward carets show local minima of the gravitational
    potential, with red carets showing minima hosting HBR. The bases
    of upward green carets show the maximum $\Phi$ isocontour of each
    HBR, bound relative to a horizontal black line delineating the
    maximum $\Phi$ of its HBP. Regions between critical points are
    represented as vertical black lines, and critical points are
    horizontal black lines where those regions intersect and merge in
    this tree diagram.}
  \label{fig:tree}
\end{figure}

\section{Results}\label{sec:results}

\subsection{Structure geography and object dendograms}\label{sec:resultmaps}

Sample surface density snapshots from the MHD\_2pc TIGRESS model can be seen in
\autoref{fig:progression}. Also shown, in left-hand panels,
is a comparison between HBR and HBP objects, projected onto the horizontal
plane.  In the right-hand panels, we similarly show projections of objects
defined by density thresholds $\nhmin = 10\pcc$ and $\nhmin = 100\pcc$.
This comparison highlights the smoother and more selective nature of energy-identified objects.

A sample dendogram of the HBP and HBR objects identified in
\autoref{fig:progression}e is shown in \autoref{fig:tree}.  The
dendrogram reveals several qualitative properties. For example, ISM
turbulence is of the order $v \sim 1-10 \kms$, so it is expected that
bound material must be found in wells with depths of $\Delta\Phi \sim
1-100 \mathrm{km^{2}/s^{2}}$. At a glance, this is apparent in
\autoref{fig:tree}. Most local minima, and most of the regions
represented by the tree do not host bound regions. The differences
between the tops of HBRs and the tops of HBPs, roughly represents the
total energy and corresponds to $v \sim 1 \kms$.

Furthermore, we can see that the merging criterion described in
\autoref{sec:mhbr} usually prefers the smallest scale isocontours at
this resolution, corresponding to critical isocontours containing only
1 local minimum. That is, no merging occurs to produce HBR in the
snapshot represented by \autoref{fig:tree}, and in general this is rare. A merged
HBR would appear as a green upward caret on a vertical line
(representing a volume) stemming above (containing) a horizontal line
(representing an intersection between isocontours of multiple local
minima). Qualitatively, this is because merging adds very little
$\Delta\Phi$ for each merge, as evidenced by short vertical lines in
\autoref{fig:tree} corresponding to $v \sim \kms$, but quickly moves
to larger length scales and higher velocity dispersion.

Note that \autoref{fig:progression} (panels e and f) shows that the
gas is mostly contained in a single large-scale region, which should
result in an overall potential well of the simulation. This is represented in
the dendrogram as the overall well shape, except for the large
isocontour at (-200 pc, 300 pc) corresponding to index 44. The densest
gas and the bound gas in the hierarchy tends to be near the bottom of
the overall well of the simulation.

\subsection{Gas Distribution and Object Properties}\label{sec:prop}

\begin{figure}
  \includegraphics[width=\columnwidth,keepaspectratio]{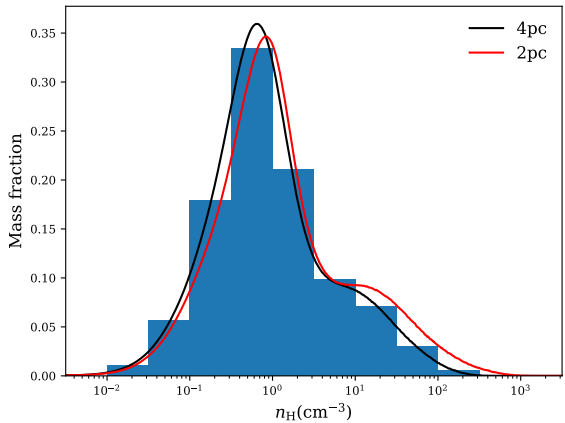}
  \caption{The hydrogen number density distributions of TIGRESS Solar
    neighborhood simulations with 2 pc (red) and 4 pc (black)
    resolution, taken at late times ($t > 300$ Myr). Half decade bins
    are shown for the 4 pc simulation case, showing the fraction of
    the total mass in each bin. }
  \label{fig:hist_bin}
\end{figure}

First, we summarize some of the basic properties of the gas in the
simulations.  In \autoref{fig:hist_bin} we show the number density
distribution in the 4pc and 2pc simulations, taken over all times. We
show mass fractions of half-decade bins in number density and
normalize the continuous distribution accordingly. In both simulations
the mass pdfs are centered near $\nh =  1~\pcc$
(mass-weighted mean densities are $\nh=4.84$ and $10.1\pcc$
for 4pc simulation and 2pc simulation, respectively)
with maximum density of
$10^{2.5}\pcc$ in the 4pc simulation and $10^{3}\pcc$ in the 2pc
simulation. The mode of the distribution is at density of $\nh=0.7\pcc$ and
$\nh=0.8\pcc$
for 4pc simulation and 2pc simulation, respectively.
The distribution depends on resolution at high number
density due to the criterion for introducing sink particles when
collapsing objects become unresolved.

\begin{figure}
  \includegraphics[width=\columnwidth,keepaspectratio]{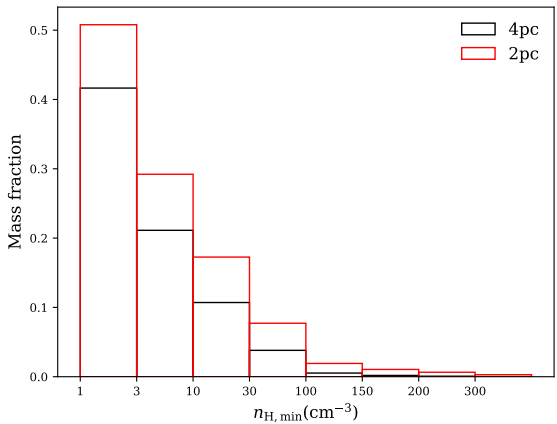}
  \caption{As in Figure~\ref{fig:hist_bin} except with cumulative
    distributions of the hydrogen number density.}
  \label{fig:hist_thresh}
\end{figure}

The density distribution is dominated by a roughly log-normal
distribution, with a secondary cold dense component. In
\autoref{fig:hist_thresh} we show the mass fractions above number
density thresholds. In the 2pc simulation, roughly half of the mass is
denser than $\nh=1 \pcc$, roughly a tenth of the mass is denser than
$30\pcc$, and a few per cent of the mass is denser than $100\pcc$.
Compared to the lower-resolution model, the 2 pc model 
has slightly larger mass fractions at higher density.

Next we describe properties of $\nhmin$ objects, using
thresholds of $\nhmin=$ 10, 30, and $100\pcc$, and compare them to HBR
objects. For the following examination of object properties we only
use the 2pc simulation so that objects are better resolved.

\begin{figure*}
  \includegraphics[width=\textwidth]{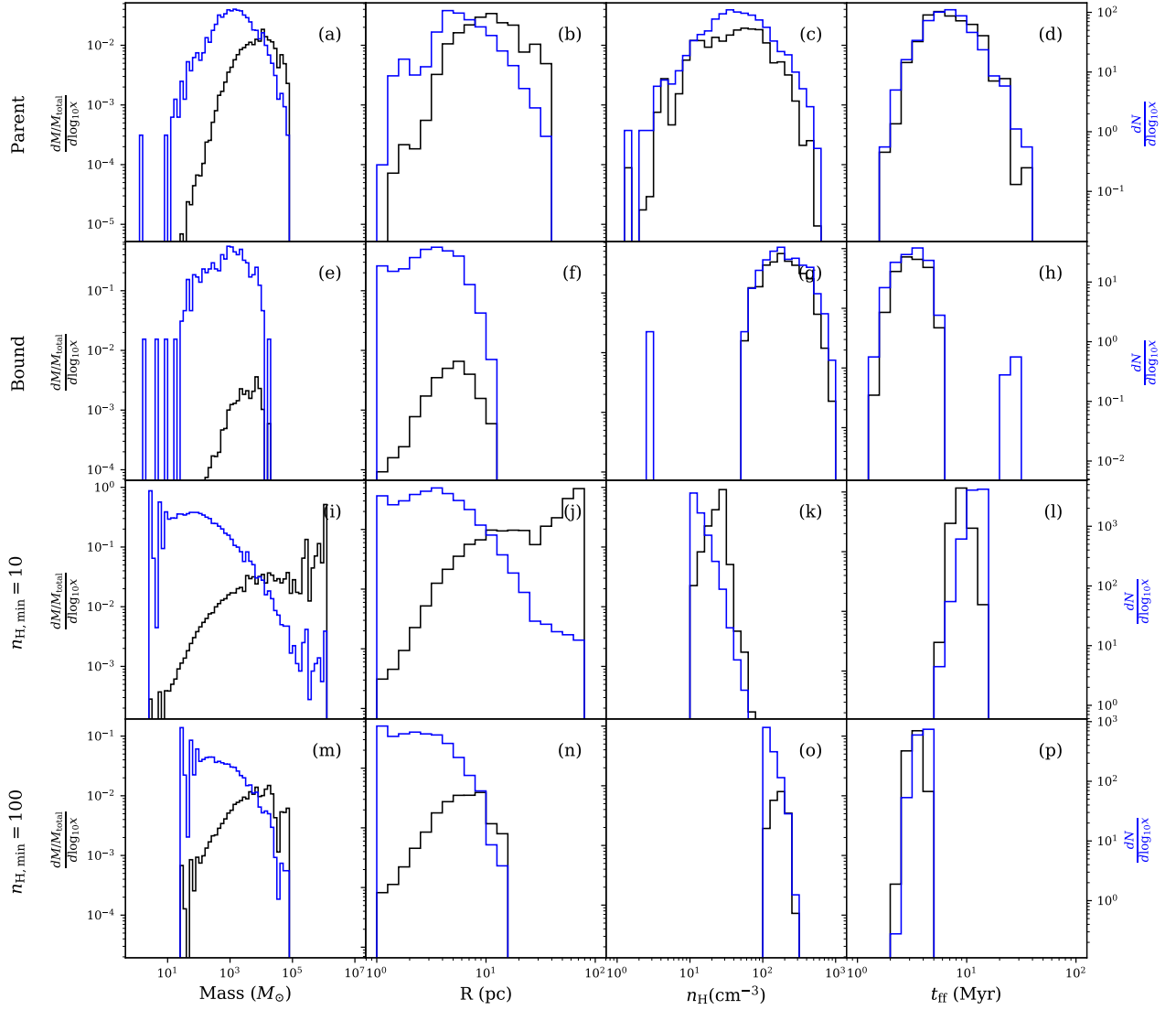}
  \caption{Number-weighted (blue, right axis) and mass-weighted
    (black, left axis) distributions of mass, radius, number density,
    and free-fall time, for objects defined in
    the 2 pc resolution simulation.  The top
    two rows show distributions for HBP (Parent) and HBR (Bound)
    objects, and the bottom two rows show distributions for objects
    defined by density threshold of $\nhmin=$ 10 and $100\pcc$. The
    radii are computed from the volume as $R = \sqrt[3]{3V/4\pi{}}$.}
  \label{fig:hist_x}
\end{figure*}

In \autoref{fig:hist_x} we present number-weighted and mass-weighted
distributions of mass, radius, density, and free-fall time for HBR and
HBP objects as well as objects defined by number density thresholds
$\nhmin=$ 10 and $100\pcc$.  For HBR objects, the typical mass is
$10^{3}-10^{4} \Msun$, with very few above $10^{4}\Msun$.
The HBR objects are mostly dense,
with $\nh$ a few $100 \pcc$.  Hence, it is useful to compare HBR to
density objects (both threshold and bin) around $100\pcc$.
The characteristic free-fall times of HBR objects are short ($\tff\sim 3$ Myr).

Radii of HBR objects are typically several pc, which demonstrates that they are
well-resolved with 2pc resolution. 
We note that a barely resolved 4x4x4 region of cells would have a
volume of $8^3=512\pc^{3}$. For number densities of 10, 30, and
$100\pcc$, such a region would respectively have masses of 170, 500,
and $1700\Msun$ assuming $\mu = 1.4$. We find very few HBR objects below this
lower mass limit set by resolution.

HBP objects have larger sizes and masses than HBR objects, with lower
characteristic densities (a few tenths) with free-fall times nearly 10
Myr.

There are a large number of $\nhmin$ objects at small
masses and radii, since we place no lower cutoff on their
size. However, most of the mass is in objects of large mass and
radius. For $\nhmin = 10\pcc$, typical (in a mass-weighted sense) objects
are $10^{5}-10^{6}\Msun$ and $\lesssim 100 \pc$. For $\nhmin = 100\pcc$,
typical objects are $10^{4}-10^{5}\Msun$ and $\lesssim 10 \pc$.

\begin{figure*}
  \includegraphics[width=\textwidth]{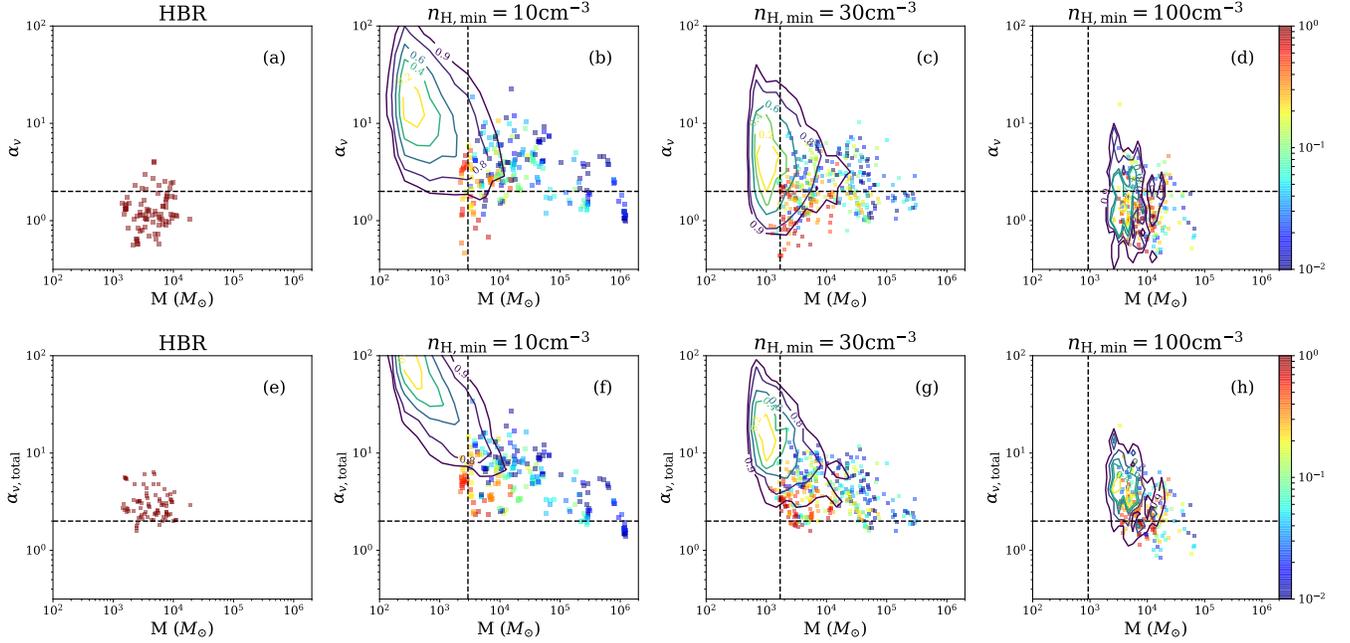}
  \caption{Distributions in virial parameters and mass of various
    defined objects in the 2pc simulation.  Left panels show HBRs, and
    central and right panels show $\nhmin = 10, 30$, and $100\pcc$
    objects. In the top row (panels a, b, c, and d), only kinetic
    energy is considered for $\alpha_{v}$ (Equation~\ref{eq:virialdef}).
    In the bottom row (panels
    e, f, g, h) $\alpha_{v,\mathrm{total}}$ considers kinetic,
    thermal, and magnetic energy (Equation~\ref{eq:virialtotdef}).
    Dashed horizontal lines delineate
    $\alpha_{v}=2$ or $\alpha_{\rm v,total} = 2$, corresponding to
    $\Esubt{k} = \Esubt{g}$ in top panels and
    $\Esubt{k} + \Esubt{B} + \Esubt{th} = \Esubt{g}$ in bottom 
    panels.  Vertical lines represent the minimum mass estimate from
    \autoref{eq:wellmass}. Contours show the distribution of
    $\nhmin$-objects whose mass has less than one per cent overlap with
    HBR. Scatter points are individual $\nhmin$-objects whose color reflects
    their mass fraction overlap with HBR. Truly bound objects (red points)
    with order unity overlap with HBR tend to be low mass (around
    $10^{3}\Msun$) with virial parameters $\alpha_v \lesssim
    2$. Especially at high masses, many apparently ``bound'' objects
    based on $\alpha_{v} < 2$ are not in fact HBR-bound (i.e. they are colored
    blue-green-yellow).  Additionally, many $\alpha_{v} > 2$ and
    $\alpha_{v,total} > 2$  objects at low and moderate density
    have significant HBR overlap (red).  These results show
    that ``observed'' virial
    parameter is not a good indication of true gravitational binding.}
  \label{fig:av}
\end{figure*}


\autoref{fig:av} shows the distribution of virial parameters and
masses for HBR and $\nhmin$ objects, via contours and scatter plots.
For scatter plots of individual $\nhmin$ objects, the color of each
point indicates the fraction of its mass that is bound, based on
overlap with HBR objects.  We find that very few $\nhmin$ objects at
the low mass end ($< 10^{3}\Msun$) have overlap with HBR, even if
their kinetic virial parameter $\alpha_{v} < 2$. $\nhmin$ objects with
negligible (less than one per cent) overlap with HBR are represented
by contours enclosing $20\%$, $40\%$, $60\%$, $80\%$, and $90\%$ of
the objects.

The depth of a well behaves as $GM/R \sim G\rho{}R^{2}$. At constant
density, only a sufficiently large object will have a well deep enough
to bind material. A rough estimate comparing $GM/R > v^{2}$ with $M =
(4\pi{}/3)\rho{}R^{3}$ yields a minimum mass that follows
\begin{equation}\label{eq:wellmass}
  M^{2} > \frac{v^{6}}{\frac{4\pi}{3}G^{3}\rho{}}.
\end{equation}
For $v = 1\kms$ and $\nhmin = 30\pcc$, this minimum mass is
$2 \times 10^{3}\Msun$. The $\nhmin = 10\pcc$ and $\nhmin = 30\pcc$ objects
overlapping HBR lie above their respective minimum mass
(for $v=1\kms$), whereas the
$\nhmin = 100\pcc$ objects all lie above the resolution minimum mass
(which is greater than the well minimum mass).

For objects of mass $\sim 10^{3.5}-10^4\Msun$, those at the  lower range of
$\alpha_v$ and $\alpha_{v,total}$ have the largest fraction of bound gas (i.e.
red colored points), which is consistent with general expectations.  However,
the actual values of $\alpha_v$ and $\alpha_{v,total}$ in the objects with
$>50\%$ HBR-overlap cover a wide range from
$\alpha_v  \sim 0.6-6$, generally decreasing at  higher density.  This shows that
the ``observed'' virial parameter is not a very accurate
quantitative measure of  gravitational boundedness.

Furthermore, Figure~\ref{fig:av} shows that unlike low-mass
($\sim 10^{3.5}-10^4\Msun$) $\nhmin$ objects with
$\alpha_v \lesssim 2$ which are generally bound, $\nhmin$ objects at the
high mass end ($ > 10^{4.5}\Msun$) have very little overlap with HBRs
even at low virial parameter ($\alpha_{v} < 2$). That is, high mass
objects ($10^{6}\Msun$) can appear bound based on simple criteria
using their mass, size, and velocity dispersion (Equation~\ref{eq:virialdef}),
but in reality this
is not consistent with full gravitational potential structure.  This
is in part due to tidal fields preferentially unbinding larger-scale
objects, and in part due to substructure. Substructure within a
$\nhmin$ object manifests itself as multiple separate HBR objects which
comprise a small fraction of the mass because most of the mass lies in
between HBRs.

In summary, we find that overdense objects with $\alpha_v \lesssim 2$ are 
truly bound only if their masses are low; high-mass objects are generally
unbound even when a simple estimate suggests otherwise.

\begin{figure}
  \includegraphics[width=\columnwidth,keepaspectratio]{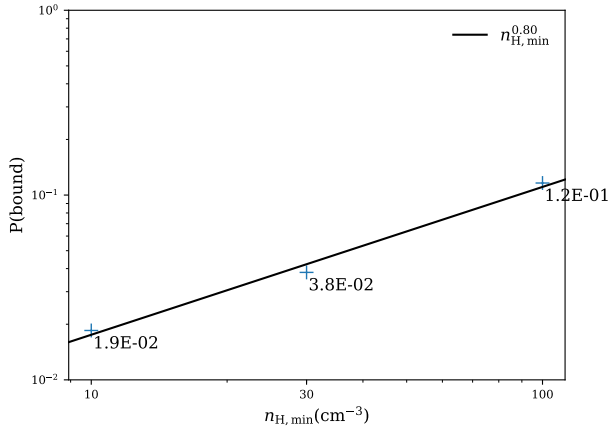}
  \caption{The fraction of mass in objects defined by density
    thresholds that is also within HBR objects; this is equivalent to
    the conditional probability $P(\rm bound | \rm dense)$, in the 2pc
    simulation. This increases with density $\nhmin$, but even at high
    densities (100 $\pcc$) only $\sim 10\%$ of the mass is bound.}
  \label{fig:pbound}
\end{figure}


In \autoref{fig:pbound} we show the HBR fraction of the mass above
number density thresholds $\nhmin = 10$, $30$, and $100
\pcc$, representing the probability of gas being bound given that it
is dense. This fraction is only a few per cent for $\nhmin = 10\pcc$ and
$30\pcc$, increasing to $10\%$ for $\nhmin = 100\pcc$. The fraction
roughly follows the power law
$\nhmin^{0.80}$. Since HBR mass tends to be
dense, each density threshold should contain nearly all the HBR
mass. Thus, the overlap fraction should roughly follow the reciprocal
of the threshold distribution as shown in \autoref{fig:hist_thresh}
(for the 2pc simulation). The threshold distribution at these number
densities follows $\nhmin^{-0.87}$, whose reciprocal has a similar slope
as expected. Although the $\nhmin = 100 \pcc$ threshold is within an
order of magnitude of the maximum density in the simulation, and the
Larson-Penston density is one of the criteria for star particle
formation, we do not find strong evidence for a critical density for
boundedness within the range $\nh = 10-100\pcc$.

\begin{figure*}
  \includegraphics[width=\textwidth]{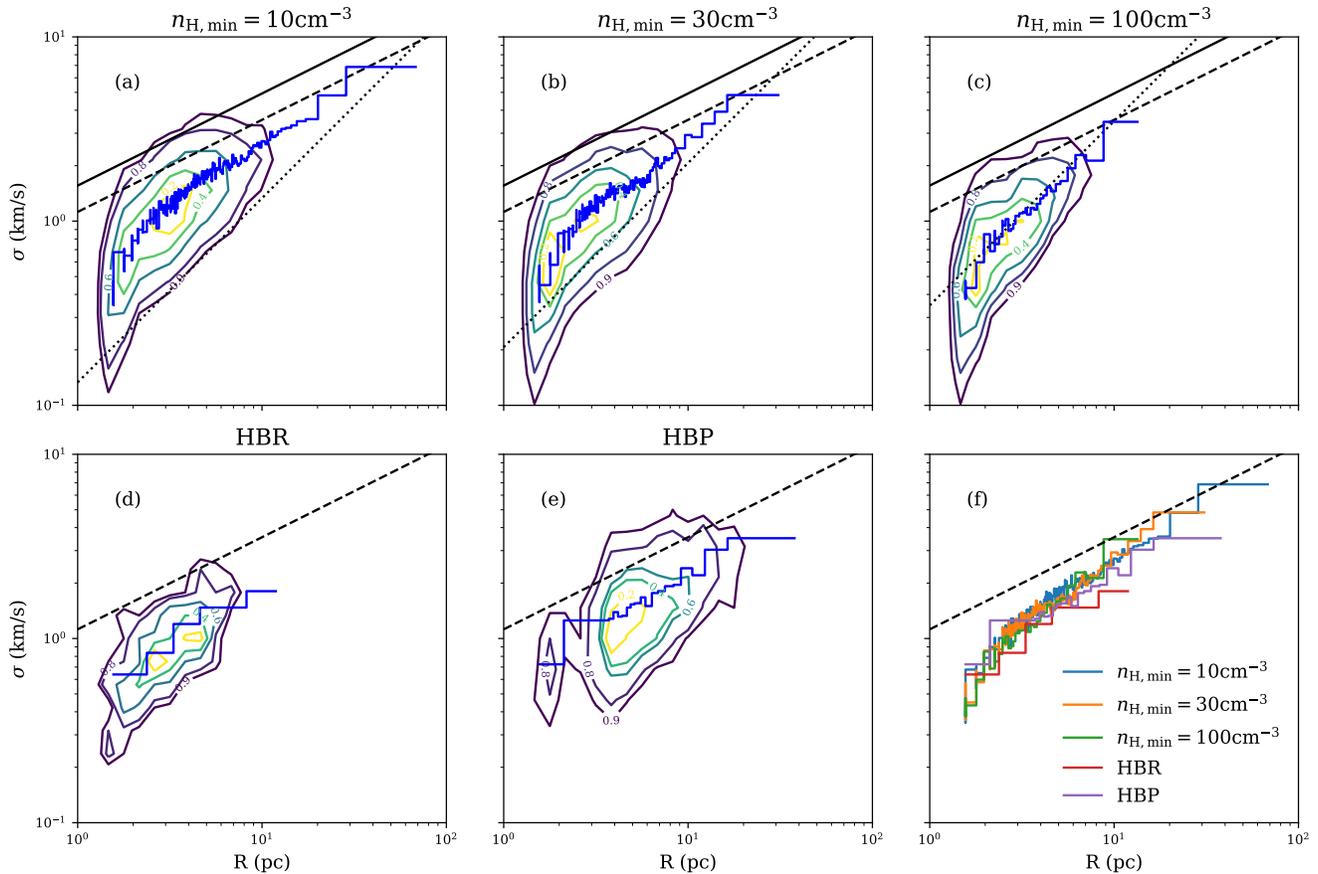}
  \caption{The linewidth-size ($\sigma_{3D}$ and radius from
    $V = (4\pi/3)R^{3}$ for object volume $V$)
    relationship of various objects (number
    density threshold, HBR, HBP) from the 2 pc TIGRESS solar
    neighborhood simulation.  In panels a-e, contours show the full
    distribution and the blue line shows the median value of radius
    bins. The contours contain $20\%$, $40\%$, $60\%$, $80\%,$ and
    $90\%$ of the objects.  For reference, the black dashed line
    represents $\sigma \propto R^{1/2}$ with normalization set by the
    measured velocity dispersion of $T < 2\times{}10^{4} K$ gas for
    $R$ equal to the measured scale height in the simulation.  In
    panels a-c, the black dotted line represents spheres at the threshold
    density with equal kinetic and potential energy
    ($\sigma \propto R \nhmin^{1/2}$).
    The solid black line is $\sigma \propto R^{1/2}$ with normalization similar
    to that of Milky Way GMCs
    $\sigma_{3D} = \sqrt{3} \times 0.9\kms (R/\pc)^{1/2}$ \citep{Heyer2015}.
    The median relations of all object
    types are stacked in the bottom-right panel.}
  \label{fig:lws}
\end{figure*}


We show in \autoref{fig:lws} the linewidth-size relationships of
$\nh$-threshold, HBR, and HBP objects.
Here, the linewidth for an object is defined as $\sqrt{2\Esubt{k}/M}$ where $M$ is the mass and $\Esubt{k}$ is the kinetic energy in the object's center of mass frame.
The size is computed from object volume $V$ taking $V = (4\pi/3)R^{3}$.

Panels a-e of the figure
include both contours of each distribution and median relationships, with
bins for the latter chosen to each contain 100 objects, except for the
final bin.  We also include several reference lines with slopes
$\sigma \propto R^{1/2}$ and $R$ for comparison.
$\nhmin$ objects and HBRs have linewidth-size relationships that are
somewhat steeper than $\sigma \propto R^{1/2}$, which is what would
be expected if all the objects simply sampled from the same power
spectrum of highly compressible ISM turbulence with an outer scale
much larger than typical object sizes.  The objects with the closest
linewidth-size relation to $\sigma \propto R^{1/2}$ are HBPs, which
are formed by isocontours of the gravitational potential.  While
there are detailed differences among the median linewidth-size
relationships for different object categories, 
\autoref{fig:lws}f shows that the median relationships are in fact quite
similar across all categories.

For $\nhmin$-objects, in each panel of \autoref{fig:lws} we also include
the line corresponding to a spherical object with density equal to the
threshold density and kinetic energy equal to the potential energy,
which has slope $\sigma \propto R \nhmin^{1/2}$.  Moving to higher $\nhmin$ shifts the
$\sigma \propto R$ line from self-gravitation upward since
$\sigma_{3D} = (\rho G 8\pi/5)^{1/2} R$.

For density threshold $\nhmin=10, 30 \pcc$, most
objects lie above the marginally self-gravitating locus.  For the
$\nhmin=100 \pcc$ threshold, the median relationship follows the
marginally-bound relation quite well.  These results  are consistent with the
results for the distributions of $\alpha_v$ shown in \autoref{fig:av},
which shows that the typical $\alpha_v$ decreases with $\nh$ threshold.
We again emphasize that even with order-unity $\alpha_v$, most of the material
in objects above a density threshold $\nhmin=100 \pcc$ is not part of bound
regions.

\subsection{Time Series}\label{sec:rtime}

We now turn to the results of our time series analysis, based on methods
described in
\autoref{sec:mtime}. Since a large number of snapshots is necessary for this
analysis,
we primarily use the 4pc simulation, which was run for a longer time.

\begin{figure}
  \includegraphics[width=\columnwidth]{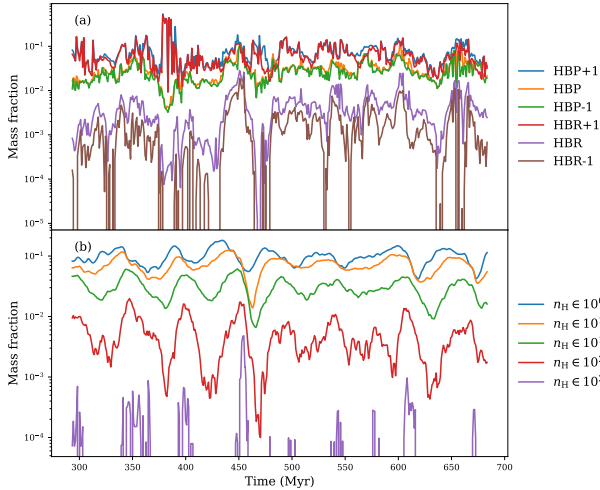}
  \caption{Time series of mass fractions for various categories of
    objects and density bins, as labeled, for the 4 pc resolution
    Solar neighborhood simulation. HBRs are bound objects and HBPs are
    their parents; see text in \autoref{sec:mhbr} for explanation of
    treatment of surface terms in HBR$\pm$1.  The quasi-periodic
    variations in the $\nh$ bin time series reflect the natural 50 Myr
    vertical oscillation timescale in the galactic potential. }
  \label{fig:time_mass}
\end{figure}

In \autoref{fig:time_mass} we show how the mass fractions of different
categories of material evolve over time.  The top panel shows HBR and HBP
material as well as HBR$\pm$1, where the latter allows for surface terms
to help confine or else disperse material, and HBP$\pm$1 are the ``parents''
of HBR$\pm$1 objects (see \autoref{sec:mhbr} for details of definitions).
HBP+1 and HBR+1 include more mass and HBP-1 and HBR-1 include less mass,
as described in \autoref{sec:mhbr}.
The bottom panel shows the mass fractions of material
in half-decade number
density bins.
Overall, the amplitude of fluctuations increases for categories with lower
mean mass fractions.  In addition, upward fluctuations are successively
delayed in time for increasing $\nh$ bins; we discuss this effect
further below.  

Roughly $10\%$ of the mass
is in low density bins and in HBR+1/HBP+1 objects. HBR+1 and HBP+1 objects
effectively subtract the surface value of kinetic, thermal, and
magnetic energy from each cell in the object, such that 
low density material in the
shallower regions of a potential well is considered bound and the object
mass is higher than if surface terms were neglected.
The HBR+1 mass is close to the
HBP+1 mass because nearly all the material within an isocontour is
considered bound.
A few per cent of the mass is within HBP-1 and HBP objects,
which have relatively similar mass histories.
Only of order $10^{-3}$ of all the material in the simulation is in 
HBR and HBR-1 objects.  Relative to HBR, HBR-1 objects have slightly less
mass because the surface energy terms are added to each cell, reducing the
amount of material that is considered bound.  

\begin{figure}
  \includegraphics[width=\columnwidth]{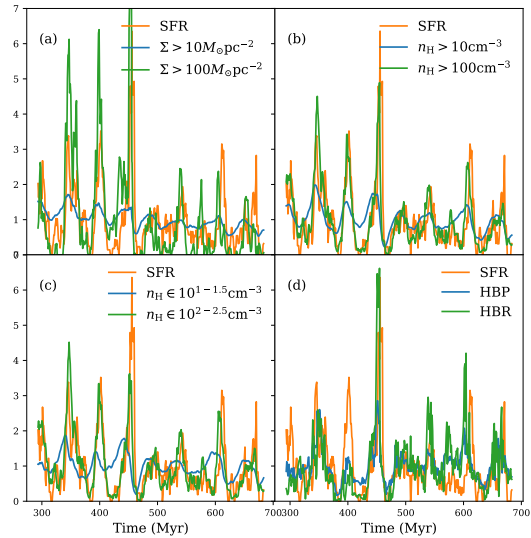}
  \caption{Comparison between star formation rate (SFR) and mass
    divided by free-fall time $M/\tff{}$ of various gas populations from
    $t=300-700$ Myr in the 4 pc resolution Solar neighborhood
    simulation. The SFR shown is smoothed over 5 Myr and normalized to
    its time-average value. The $M/\tff{}$ time series are also
    normalized relative to their time-averaged values. Individual
    panels compare (a) surface density thresholds, (b) number density
    thresholds, (c) number density bins, and (d) HBP and HBR. In
    all cases the denser, more restrictive definition leads to a
    qualitatively better match between SFR and $M/\tff{}$.}
  \label{fig:mbtsfr}
\end{figure}

It is interesting to compare the SFR history with the evolution of
$M/\tff$ at lower and higher gas surface density $\Sigma$, lower and
higher density $\nh$, and for less and more bound objects.  By
comparing time series in \autoref{fig:mbtsfr}, it is evident that
more restrictive definitions have an improved correlation with $\SFR$.
This holds
for increasing threshold $\Sigma$, increasing $\nh$ threshold,
increasing $\nh$ bins, and increasing boundedness from HBP to HBR.
Intriguingly, \autoref{fig:mbtsfr}b,c demonstrate that simple density
criteria (high density threshold or bin) yield {\it better}
correlation with SFR than the more complex criteria based on total
energy in the full gravitational potential landscape that go into the
definition of HBR, shown in \autoref{fig:mbtsfr}d. The visual
impressions of these histories already suggest that gravitational
binding is not a guarantee that star formation will be successful; we
return to this quantitatively below.

\begin{figure*}
  \includegraphics[width=\textwidth,height=0.9\textheight,keepaspectratio]{\figtimedensity}
  \caption{Comparison between SFR and $M/\tff{}$ as in
    \autoref{fig:mbtsfr} for half-decade number density bins
    Here
    $M/\tff{}$ is normalized by a factor of $\eff{}/\mean{\SFR}$,
    where $\eff{}$ is the result of the simple linear regression
    (\autoref{eq:effinfer}).  The inferred delay time $t_{d}$ is used
    to offset the time series, which is labeled as ``delayed'' for
    each number density bin. It is clear that the correlation between
    SFR and $M/\tff{}$ improves and $t_{d}$ decreases as density
    increases.}
  \label{fig:time_density}
\end{figure*}

\begin{figure}
  \includegraphics[width=\columnwidth,keepaspectratio]{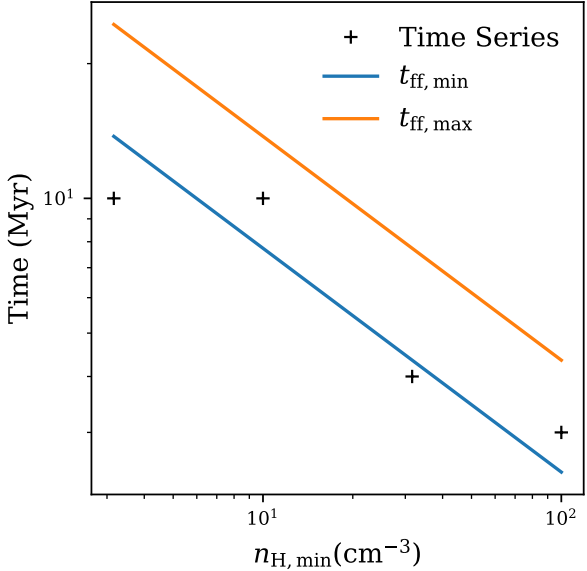}
  \caption{The delay time $t_{d}$ vs. lower density edge of some of
    the half-decade density bins shown in
    \autoref{fig:time_density}. Lines for
    $\tff{} = [3\pi/(32G\mu n_{\mathrm{H}})]^{1/2}$
    are computed using the
    lower and upper number density edge, respectively resulting in a
    maximum and minimum free-fall time. For denser bins, the delay
    time roughly follows the minimum free-fall time.}
  \label{fig:time_lag}
\end{figure}

\begin{figure*}
  \includegraphics[width=\textwidth,height=0.9\textheight,keepaspectratio]{\figtimeothers}
  \caption{Comparison of $\SFR$ and $\epsilon_{\rm ff} M/\tff$ as
    in \autoref{fig:time_density} for surface density thresholds
    ($\Sigma=$ 10 and $30\Msun\pc^{-2}$), number density thresholds ($\nhmin=$ 10, 30, and $100\pcc$), and HBR (bound) objects, showing only the delayed time
    series.}
  \label{fig:time_others}
\end{figure*}

As described in Section \ref{sec:mtime}, we can apply linear
regression to obtain both the optimal time delay to match the shape of
each $M/\tff$ time series to the SFR, and the corresponding
normalization amplitude $\eff$ that measures the best-fit star
formation efficiency per free-fall time. \autoref{fig:time_density}
shows the result of applying this linear regression, demonstrating
that higher $\nh$ bins correlate more strongly to SFR, with a smaller
time delay, compared to lower $\nh$ bins.  Even at low
$\nh = 10^{0.5}\pcc$ some correlation is apparent, but not for lower
densities. Thus, while the amount of gas in low density bins
($\nh=10^{-0.5}-10^{0.5}$) comprising the bulk of ISM mass (see
\autoref{fig:hist_bin}) varies in time due to large-scale vertical and
horizontal oscillations that produce ISM compressions and
rarefactions, these variations do not appear to directly {\it induce}
star formation.

In \autoref{fig:time_lag} we show that the time delay inferred from
the fit is comparable to the free-fall time associated with the $\nh$
bin upper edge,
$t_{\mathrm{ff,min}} = [3\pi/(32G\mu n_{\mathrm{H,max}})]^{1/2}$, with
$n_{\mathrm{H,max}} = 10^{0.5}n_{\mathrm{H,min}}$. This is consistent
with the idea that a variation in mass or $M/\tff$ at a given density
can only lead to a variation in SFR after the gas is able to dynamically
respond; the minimum response time is the free-fall collapse  time at that
density, and this indeed appears to be the defining timescale, even
though the efficiency of collapse is less than unity.

In \autoref{fig:time_others} we compare histories of star formation
with the time series of $\epsilon_{\rm ff} M/\tff$ for best-fit time
delay and $\epsilon_{\rm ff}$ for material defined by surface density
threshold, number density threshold, and HBR objects. For surface
density and number density, correlation with SFR improves with a
higher threshold. This correlation is visually similar for $\Sigma >
30 \surfunit$, $\nh > 10, 30\pcc$.  Although HBR and $\nh > 100\pcc$ both
follow $\SFR$ quite closely (and in particular follow lows much better than predictions based on lower density thresholds),
in fact the mass of HBR gas provides a slightly worse prediction of $\SFR$
than the mass of gas at $\nh > 100\pcc$.  We quantify this in the next
subsection.

\subsection{Star Formation Efficiency}\label{sec:eff}

\begin{figure*}
  \includegraphics[width=\textwidth,keepaspectratio]{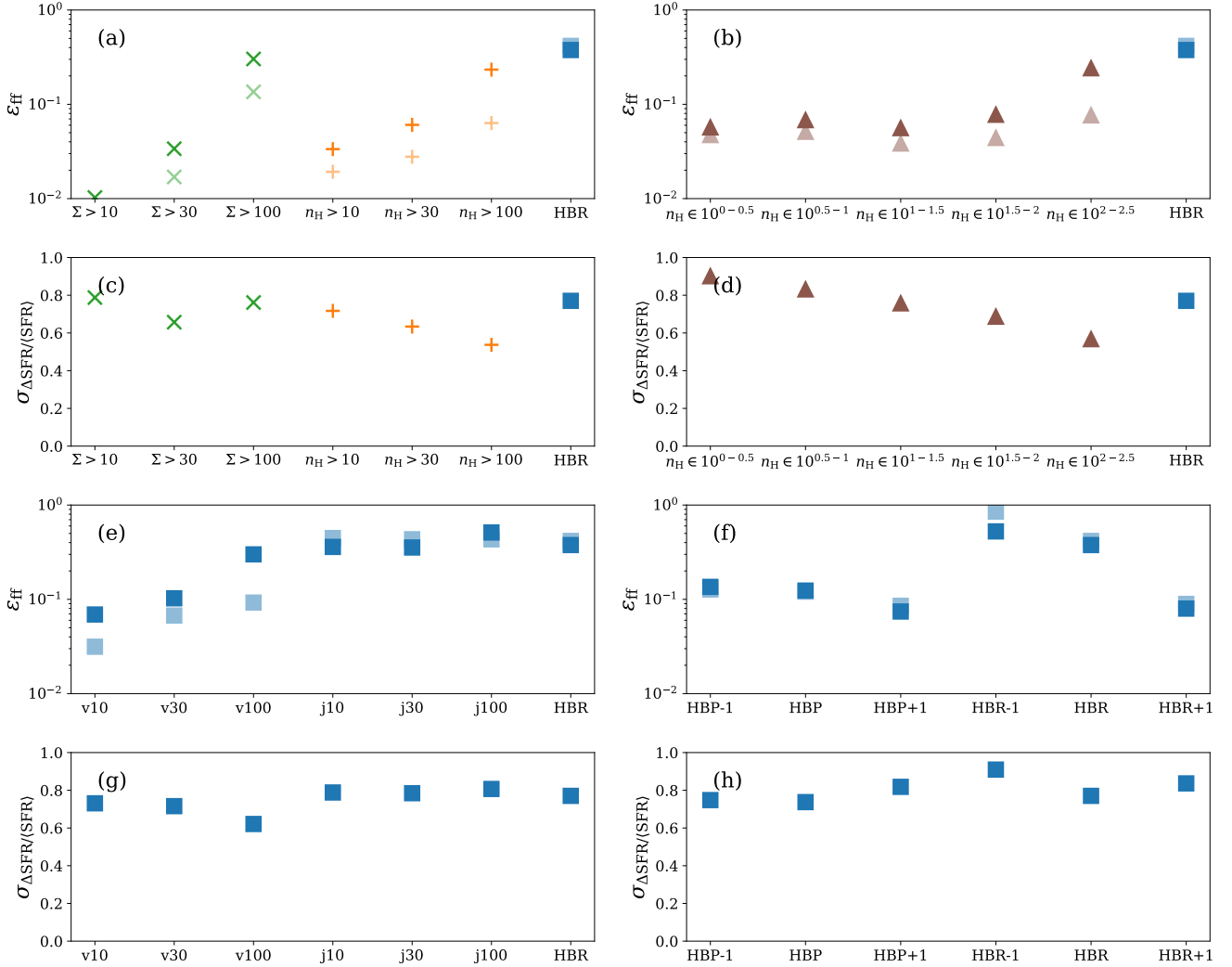}
  \caption{Inferred $\eff{}$ (a, b, e, f) and RMS error $\seff$ (c, d
    , g, h) based on time series of SFR compared to time series of
    $M/\tff$ for selected subsets of the gas, as labeled.  Gas
    selection criteria include density ($\nh$) thresholds and bins,
    surface density ($\Sigma$) thresholds, bound objects (HBR) and
    their parents (HBP), overlaps between density and HBR objects
    ($j$), combined density and $\alpha_v<2$ criteria ($v$); see text
    for details.  In each panel, results from 4pc are shown as darker
    points with 2pc ($\eff$ only)shown as lighter points. Errorbars from \autoref{eq:Bayesvariance} are not shown but would lie within markers, decreasing with the number of time snapshots used.}
  \label{fig:eff}
\end{figure*}

In this section, we present results on our inference for efficiency
per free-fall time $\eff$ for various subsets of gas, based on
application of the methods of Section \ref{sec:mtime}.  We are interested
in both the measures of $\eff$ and how they depend on the criteria for
defining a subset of the gas, and quantitative assessment of the
relative performance for predicting SFR.  We include results from both
the 4pc simulation time series, which based on its larger number of
snapshots is advantageous in terms of sample size, and the 2pc
simulation, which allows us to test whether our results are converged
with respect to numerical resolution.

\autoref{fig:eff} shows for several different categories of objects
the values for $\eff$ and for $\sigma_{\Delta \SFR/\langle \SFR \rangle}$ based on linear
regression.  The top two rows show results for gas subsets defined by
$\Sigma$ and $\nh$ thresholds and by $\nh$ bins, also comparing to HBR
results.  For both $\Sigma$ and $\nh$ thresholds
(\autoref{fig:eff}a,c), the total mass decreases faster than the
free-fall time as the threshold increases.  As a result, $\mff$ decreases at
increasing density,
leading to an increase in $\eff$ with threshold level.  At the same time,
$\sigma_{\Delta \SFR/\langle \SFR \rangle}$ 
mostly decreases with increasing threshold, implying
better correlation of denser gas with SFR; this is consistent with the
visual impression from previous plots.  
When we consider subsets of gas in density bins (\autoref{fig:eff}b,d),
$\eff$ increases and $\sigma_{\Delta \SFR/\langle \SFR \rangle}$
decreases at higher densities $\nh$.
However, at $\nh < 10^{1.5}\pcc$, $\eff$ is roughly flat at roughly $0.06$.
The value $\eff \sim 0.01$ for $\Sigma > 10 \surfunit$ represents the mean 
efficiency for the bulk of the material in the simulation.

Interestingly, while the
values of $\eff$ for high density thresholds are similar to the value
of $\eff$ for gas in HBR (a few tenths), the value of
$\sigma_{\Delta \SFR/\langle \SFR \rangle}$ 
is {\it lower} for density thresholds than for HBR gas.  While HBR gas is
mostly quite dense, this says that the additional criteria of
requiring that every parcel of gas is bound within an HBR does
{\it not} lead to better agreement in the histories.

Looking at HBP and HBR variants in \autoref{fig:eff}f,g, higher mass
variants (all HBP, and HBR+1) have lower $\eff$ ($\sim 0.1$).
 HBR has $\eff \approx 0.4$ and HBR-1 has
$\eff \approx 0.6$: in both cases the efficiency is nearly order
unity, as might be expected of truly collapsing objects.  However,
we cannot distinguish between the case that each bound object takes a total time
$\tff/\eff$ to collapse, vs. the case that the probability
of each bound object being dispersed is $1-\eff$ and the collapse
timescale for surviving objects is $\tff$. 
The variants of HBR and HBP have a similar correlation to SFR, but HBR
and HBP have slightly lower $\sigma_{\Delta \SFR/\langle \SFR \rangle}$.
It is interesting that isocontours alone (HBP) provide a
reasonable correlation to SFR, and that varying treatment of surface
terms has little effect.

We can also consider combined criteria and test the correlation with
SFR.  In \autoref{fig:eff}e,g, we show results for ``$j$'' time
series, consisting of material that exceeds certain density
thresholds and also overlaps with HBRs, and for ``$v$'' time series,
material in $\nhmin$ objects that also satisfy $\alpha_{v} < 2$
(kinetic energy only, excluding thermal and magnetic, in the virial
parameter).
The $v$ series
has similar results to $\nhmin$ objects themselves, but with
slightly greater $\eff$ and comparable
$\sigma_{\Delta \SFR/\langle \SFR \rangle}$.    
Whereas $\eff$
for $\nh > 10\pcc$ doubles when considering $\alpha_{v}$ in the $v$ series,
it is only higher by $30\%$ in the $\nh > 100\pc$ case.
Considering the virial parameter mainly affects lower density gas.

The ``$j$'' series over material overlapping between HBR and $\nhmin$
objects is mostly similar to HBR because most HBR mass also
satisfies $\nh \sim 100\pcc$. The exception is $\nh > 100\pc$ in the
4pc case because $100\pcc$ is close to the maximum density of the
simulation.

Note that although $\eff$ for surface and $\nhmin$ objects are
resolution dependent (4pc and 2pc points differ), the values of $\eff$
for low- and moderate-$\nh$ bins are resolution independent.  Thus,
differences in $\nhmin$ objects are due to the lack of gas in high
density bins for lower resolution simulations.  Unlike the high-$\nh$
series, the HBR and HBP series (and their variants) have essentially
the same values of $\eff$ at 4pc and 2pc resolution.  The difference
in $\sigma_{\Delta \SFR/\langle \SFR \rangle}$ with resolution primarily reflects the different number of snapshots available at each resolution.
Whereas multiple cycles of star formation are available in the 4pc simulation, only 70 Myr of snapshots are available from the 2pc simulation. Therefore, we do not draw any conclusions from the 2pc values of $\sigma_{\Delta \SFR/\langle \SFR \rangle}$ and do not show those values in \autoref{fig:eff}.

\begin{figure}
  \includegraphics[width=\columnwidth,height=0.9\textheight,keepaspectratio]{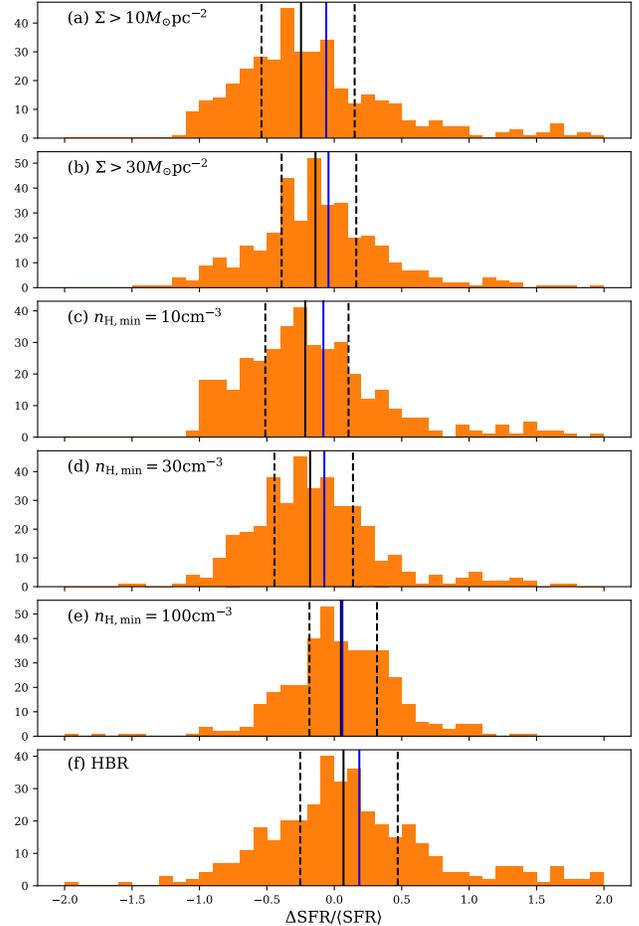}
  \caption{A comparison of the distribution of $\Delta\mathrm{SFR}/\langle\mathrm{SFR}\rangle$ (see Equation~\ref{eq:deltaSFR}), the difference between actual and predicted SFR using different categories of gas, based on (a)-(b) gas surface density threshold, (c)-(e) number density threshold, and (f) HBR objects,
    corresponding to the time series shown in \autoref{fig:time_others}.
    The mean error is shown as a blue vertical line, and quartiles are shown in black lines (median in solid, 25 and 75 dashed).}
  \label{fig:error_hist}
\end{figure}

In \autoref{fig:error_hist} we illustrate how correlation changes for
different gas selection criterion
in further detail by providing histograms of the
error $\Delta\mathrm{SFR}/\langle\mathrm{SFR}\rangle$
(Equation~\ref{eq:deltaSFR}), where a
positive error means that the simulated star formation rate is higher
than the model star formation rate based on $M/\tff$ for a single snapshot. 
For less restrictive selection criteria, such as lower $\nh$ or $\Sigma$,
the mean and median in the distributions shift to the left, indicating that the
predicted $\eff M/\tff$ exceeds the actual SFR.  This is clearly  evident in
\autoref{fig:time_others}: for low density thresholds, there is little
predicted variation in the SFR about the mean, whereas the true SF history is
mostly below the mean level, with some sharp peaks.   
\autoref{fig:error_hist} also shows that worse correlation at lower $\nh$ or
$\Sigma$ threshold is associated with  
a larger number of snapshots wherein $\Delta\mathrm{SFR}/\langle\mathrm{SFR}\rangle \sim -0.5-1$
and $\Delta\mathrm{SFR}/\langle\mathrm{SFR}\rangle > 1$.
Again, this is evident in the missed
``long valleys'' and ``sharp peaks''
for the prediction based on $\eff M/\tff$ in the
$\Sigma > 10 \surfunit$, $\nh > 10 \pcc$ cases in
\autoref{fig:time_others}.
Since missing sharp peaks occurs during periods of high SFR, considering an alternative version of \autoref{eq:SFRerr} by weighting by $\SFR(t_{i})$ would amplify the improvement of correlation with increasing density. 
\autoref{fig:error_hist}e,f also quantifies the visual impression from \autoref{fig:time_others} that
the restriction that gas be bound (HBR) does not offer better predictive
power for $\SFR$ compared to a simple high density threshold. In particular,
the HBR prediction misses a peak at $t\sim 400$Myr, which accounts for the
positive error tail in $\Delta \SFR$ compared to $\nh  > 100 \pcc$.

Counterintuitive to the immediate
visual impression from \autoref{fig:time_others},
HBR has larger $\sigma_{\Delta \SFR/\langle \SFR \rangle}$ than even the $\nh >10 \pcc$ and
$\nh > 30 \pcc$ for the 4pc model, although for the 2pc model HBR performs
better.  The primary reason for this is the overall much larger range of
predicted $\SFR$ from HBR; since this has high peaks that can be slightly
offset from the peaks in the true $\SFR$, this leads to a broader distribution
of errors in \autoref{fig:error_hist}.
Another reason is that HBR can be too selective, where there are snapshots with high SFR but insufficient corresponding HBR gas mass.

\subsubsection{Dependence on virial parameter}

\begin{figure}
  \includegraphics[width=\columnwidth,height=0.9\textheight,keepaspectratio]{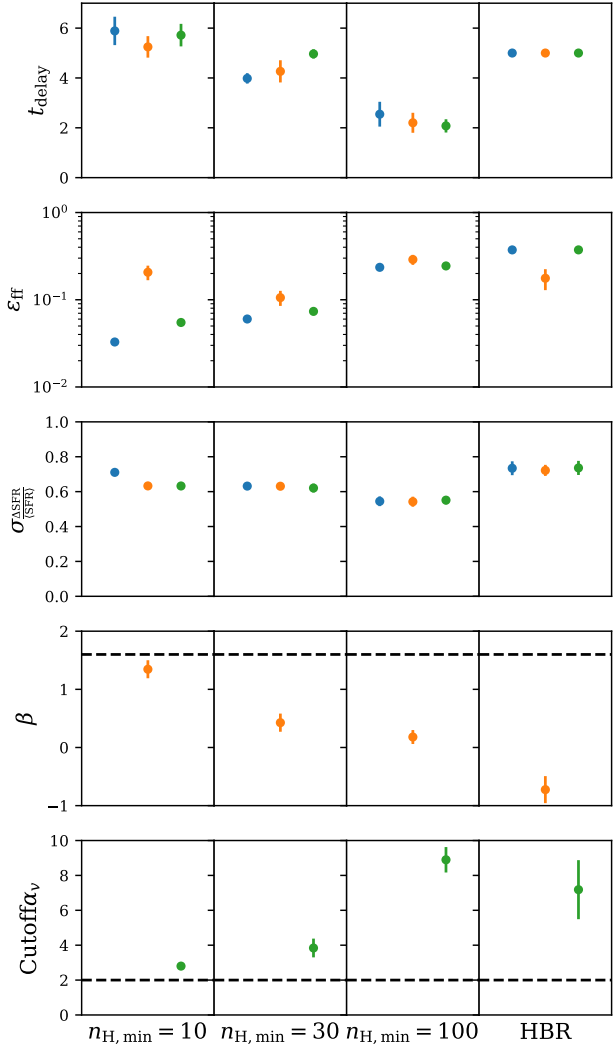}
  \caption{Comparison of inferred model parameters and goodness of fit for three
    models of the dependence of SFR on $\alpha_v$ as described in
    Section \ref{sec:effmodels}.    Results are shown for model with
    no $\alpha_v$ dependence (blue points), model with exponential
    dependence on $\alpha_v^{1/2}$ (orange points;
    \autoref{eq:padoanmodel}), and model with an $\alpha_v$ cutoff
    (green points; \autoref{eq:cutoffmodel}). Points and bars represent the mean
    (\autoref{eq:Bayesmean}) and standard
    deviation (from \autoref{eq:Bayesvariance})
    of marginalized distributions for time delay
    $t_{\mathrm{delay}}$ (in Myr), efficiency $\eff$, slope $\beta$ for the
    exponential model, and cutoff $\alpha_{v}$.
    Standard deviations of normalized $\SFR$ errors
    $\sigma_{\Delta\SFR/\langle\SFR\rangle}$
    (inferred $\sigma$ in \autoref{eq:likely}) are shown for all models.
    Reference values $\beta = 1.6$ and $\alpha_{v} = 2$ are shown with
    horizontal dashed lines.  Columns left to right use thresholds
    $\nh>10$, $30$, and $100 \pcc$, and energy-based
    criteria (HBR) to define objects.}
  \label{fig:cphne}
\end{figure}

As discussed in Section \ref{sec:effmodels}, we can apply Bayesian
inference to our time series to evaluate parameters and explore the
relative goodness of fit for different models that have been proposed
for the dependence of star formation on the virial parameter.  In our
tests, we separately examine objects defined by number density
thresholds with $n_{\mathrm{H,min}} = 10, 30$, and $100\pcc$, as well
as HBR objects.

\autoref{fig:cphne} presents the results of our analyses.  From left
to right, panels show results for objects defined by different density
thresholds and by the HBR criterion.  Each row gives values of
parameters obtained for the three models under consideration: constant
$\eff$ (blue points, equivalent to the results presented in
\autoref{fig:eff}), an exponential dependence on $\alpha_v^{1/2}$
(orange points, generalizing \citealt{2012ApJ...759L..27P}), and a
cutoff in $\alpha_v$ (green points).
The plotted value of $\eff$ represents
$\epsilon_{\rm ff,0}$ for the exponential and cutoff models.

For all models, from lower density to higher density thresholds the
inferred time delay decreases, consistent with
\autoref{fig:time_density}.  At the same time, the inferred $\eff$ 
increases with $\nh$ for both the constant-$\eff$ and the
virial-cutoff models.
The inferred $\epsilon_{\rm ff,0}$ does not
monotonically vary with  $\nh$ for the exponential model.  
Going from $\nh >30$ to $\nh >100$, the RMS error $\sigma_{\Delta\SFR/\langle\SFR\rangle}$
decreases for both the exponential and cutoff model, similar to what was
shown previously for the model with no $\alpha_v$ dependence.  Based on 
RMS error levels for any given gas selection criterion, there is generally
no significant preference for the $\alpha_v$-dependent models for $\SFR$ compared to the
$\alpha_v$-independent model.   This can be understood considering the 
inexact correspondence between apparent $\alpha_v$ and true boundedness,
and the previously-discussed limitations of boundedness as a detailed
predictor of the $\SFR$.
The exception is the lower-density
$\nh>10$ object class, in which both models that
account for the virial parameter perform better than the
constant-$\eff$ model (see further discussion below).  

For the exponential model, only the low density
threshold case shows a similar slope $\beta \approx 1.6$ to that found by 
\citealt{2012ApJ...759L..27P}; at high density
$\beta \sim 0$ is preferred, which is equivalent to constant $\eff$.
This is consistent with expectations, considering the differences between
the types of simulations of \citealt{2012ApJ...759L..27P} vs.
the current models and analysis.  In \citealt{2012ApJ...759L..27P}, the
comparison was between global SFRs for small-box simulations of cold gas
in which turbulence was driven to different levels.  This is most similar
to our selection of objects with $\nh > 10 \pcc$, and then assigning 
a relative probability of SF depending on level of the virial parameter
(which has a large variation at low density, with only the lower $\alpha_v$
objects being massive enough to host star formation,
as shown in \autoref{fig:av}).
When we instead select objects that are already quite overdense compared
to the average, the range of virial parameters is smaller
(as seen in \autoref{fig:av}) so there is little ``leverage.''

For a low density threshold $\nh > 10\pcc$, the $\alpha_{v}$ cutoff
model prefers $\alpha_{v} \approx 2$, for similar reasons to the higher
preferred $\beta$ in the exponential model.  However, at $\nh>30$ and
$\nh>100$, the inferred cutoff $\alpha_v$ values are larger, demonstrating
that the density threshold itself provides a good correlation with the
SFR and that removing high $\alpha_{v}$ material is not
preferred. Energy-selected HBR objects also do not benefit from a
virial parameter cutoff.

\section{Conclusion}\label{sec:conc}

\subsection{Summary}\label{sec:summary}

In this work, we have applied structure-finding techniques to
TIGRESS simulations of the star-forming ISM, and characterized the
properties of the objects we identify.  In addition, we have
investigated families of relationships between the SFR and material
that could be considered ``eligible'' for star formation, by being
part of a subset of the gas with defined properties.  For the latter,
we consider both collections of objects and more general gas
subsets.

Our primary comparison of structures is between those  
defined based on density or surface density (bins or thresholds)
and those that are defined based on the gravitational potential (also
considering kinetic, thermal, and magnetic energy).  The former is
more analogous to the definitions of ISM structure typically used in
observations (where boundaries are often defined by observed
intensity), whereas the latter more directly connects to dynamics.
The definitions and techniques used to identify structures are
described in Section \ref{sec:mhbr} and \ref{sec:mprop}.

For both material defined by density selection criteria and material
defined by energy selection criteria, we compare time series of $M/\tff$
to the SFR history.  We use these comparisons to fit for time delays ($t_d$)
and 
efficiencies per free-fall time ($\eff$).  In addition, we apply Bayesian
inference to compare three different models for star formation with different
dependence on the virial parameter.  Techniques are described in
Section \ref{sec:mtime} and \ref{sec:effmodels}.

Key results are as follows:

1. {\it Object properties}:
Basic statistics
(mass, size, density, free-fall time) of HBRs (bound
objects) and HBPs (their parents) are compared to statistics of $\nhmin$
objects defined by density contours in \autoref{fig:hist_x}.  Typical masses
of HBRs are $\sim 10^3-10^4 M_\odot$, with $\nh \sim 100  \pcc$.  HBPs have
slightly lower density and masses that extend up to $\sim 10^5 M_\odot$.  Thus,
bound objects are dense. Most of the mass of $\nhmin$ objects is in
large structures: typical values  are
$R\sim 30-100 \pc$ and $M\sim 10^{5}-10^{6}\Msun$ for
$\nhmin = 10\pcc$, and $R\sim 10 \pc$ and $M\sim 10^{4}\Msun$ for $\nhmin = 100\pcc$.
Not all dense objects are bound (see below).

2. {\it Virial parameters and boundedness}: Figure \ref{fig:av} shows
the distribution of values of the virial parameter $\alpha_v$ most
commonly used in observations, which compares kinetic energy with
gravitational energy, assuming an isolated sphere with the same
mass and volume to compute $\Esubt{g}$
(\autoref{eq:virialdef}).  Figure \ref{fig:av} also shows results for 
a variant $\alpha_{\rm v,tot}$ that includes thermal and
magnetic energy (\autoref{eq:virialtotdef}).  Because magnetic and
kinetic energy are comparable, we find that neglect of magnetic energy
in estimating the virial parameter is not justified.  Interestingly, while HBR
objects are defined based on bound material, they have a range of
values for ``observed''
$\alpha_v \sim 0.5-5$ and $\alpha_{\rm v,total}\sim 2-7$.
\autoref{fig:av} also depicts the fraction of gas in each
$\nh$-defined object that is truly bound when considering the full
gravitational potential.  Many objects that appear bound based on
$\alpha_v$ in fact contain only a small fraction of bound gas; this is
especially an issue at high mass $M\sim 10^4-10^6 M_\odot$.  Massive,
moderate-density objects exist but they are not bound by gravitational
wells even when $\alpha_v <2$ (\autoref{fig:av}b,c).  The probability
of gas being bound increases with $\nhmin$ (\autoref{fig:pbound}).

3. {\it Linewidth-size relations}: \autoref{fig:lws} shows that the
median linewidth-size relation for low-$\nhmin$ structures is shallower
than and lies above the $\sigma \propto R$ relation for bound objects
with fixed density,
but is fairly close to the mean $\sigma \propto R^{1/2}$
relation expected for supersonic turbulent gas with
outer scale exceeding the cloud scale.
At high $\nhmin$, the median linewidth does follow the $\alpha_{v} = 2$
linewidth-size relation
$\sigma_{3D} \approx (\rho G 8\pi/5)^{1/2} R$
for $\rho = \mu \nhmin$.
HBPs follow the same
$\sigma \propto R^{1/2}$ relation as objects selected with a low
density threshold, and in both cases the
normalization is consistent with the large-scale velocity dispersion
and overall scale height of the ISM in the simulation.  

4. {\it Temporal histories}:
From the time series, we find that on average only a few tenths of
percent of the simulation mass is in bound structures (HBRs), while
$\sim 10\%$  is at densities at least an order of magnitude
above the median density ($\nh \approx 1 \pcc$) in the simulation
(\autoref{fig:time_mass}).  The time series of the bound mass also has
high variability on short (few  Myr)
timescales and large-amplitude fluctuations.
Fluctuations in the mass of gas at high densities $\nh > 100 \pcc$
exceed an order of magnitude, and the same is true for the gas mass at
high $\Sigma > 100 \surfunit$ (\autoref{fig:mbtsfr}).  In contrast,
the mass of moderate-density gas fluctuates only over a factor $\sim
3$ with a timescale comparable to large-scale galactic vertical and
horizontal oscillation times in the galactic potential.  Generally,
upward fluctuations in any mass bin are delayed relative to those in
lower-density mass bins, and star formation fluctuations are delayed
by $\sim \tff(\nh)$ relative to the mass of gas with density $\sim
\nh$ (\autoref{fig:time_density}, \autoref{fig:time_lag}).

5. {\it Star formation efficiency per free-fall time}: By correlating
the time history of $M/\tff$ in different gas subsets with the time
history of the SFR, we measure $\eff$; results are reported in
\autoref{fig:eff}.  While $\eff$ is fairly flat in density bins at
$\nh \lesssim 30\pcc$, it increases to a few tenths when $\nh >
100\pcc$. This is close to the value for bound objects ($\eff =0.4$
for HBR gas). The degree of correlation between the detailed temporal history of
$\eff M/\tff$ and SFR($t$) secularly increases with increasing
density, as shown in \autoref{fig:time_others} and \autoref{fig:eff}d.
Even though the time series of $\eff M/\tff$ for HBR gas mostly tracks
SFR($t$) quite closely (\autoref{fig:time_others}), the RMS
error (defining $\Delta \SFR = \SFR - \eff M/\tff $) is worse than for
moderate-density gas because the large-amplitude variations in the
mass of HBR gas imply any ``miss'' is strongly penalized.

6. {\it Dependence of star formation on virial parameter}: In addition
to considering the simplest star formation model prescription in which
$\eff$ is constant for all gas in a given density bin, we test two
models in which $\eff$ depends on the virial parameter $\alpha_v$ of
defined density structures.  For one model, $\eff$ has an exponential
dependence on $\alpha_v^{1/2}$, and for the other, $\eff$ is zero
above some cutoff in $\alpha_v$.  We use Bayesian inference to obtain
marginalized model parameters and RMS errors, as shown in
\autoref{fig:cphne}.  We find that allowing for a dependence on
$\alpha_v$ improves the correlation with SFR for moderate-density gas
($\nhmin = 10 \pcc$), but does not alter the strength of the correlation
for high-density gas ($\nhmin = 30,\ 100 \pcc$) or for energy-selected
HBR objects.  
Overall, we conclude that the performance and parameters for
$\alpha_v$-dependent models of star formation, when applied to the
full multiphase ISM, may depend on how objects are defined (e.g. a
dependence on density contrast relative to ambient), and/or on global
aspects of ISM dynamics and star formation (including the space-time
correlations of feedback with gas structures).

\subsection{Discussion}\label{sec:discussion}

\subsubsection{Quantifying the role of self-gravity: are GMCs bound?}

There are a number of reasons why apparent virial parameters disagree
with detailed measurements of boundedness.
For example, $\alpha_{v}$ or $\alpha_{\rm v, total}$ could
underestimate boundedness because a uniform cloud is assumed, but the
actual gravitational potential can more strongly bind material in the
center of an object if it is stratified.  Also, our HBR definition
considers gravitational energy relative to a surrounding potential
isocontour, where the potential considers all material rather than just
an isolated structure. 
Material in and beyond the HBP  surrounding an HBR contributes to
defining the bounding equipotential and to determining how
deep the potential well is. 
Thus, an HBR can be more bound than it would appear from using just 
an object's own mass in 
$\alpha_{v}$ or $\alpha_{\rm v, total}$ (as in e.g. \autoref{fig:av}e)
because mass outside of itself contributes to defining the
equipotentials and containing the gas in a local region.

At the same time, objects can also be less bound than would be
predicted based on the traditional virial ratio of \autoref{eq:virialdef},
because the assumption of an isolated object
with vacuum boundary conditions overestimates $|\Esubt{g}|$ compared to the
real case in which tidal forces limit the region that can be bound to
a given center.  Considering the gravitational potential computed
globally, including tidal forces, means that dense objects that are
near other dense objects will be less bound than the naive estimates
used in $\alpha_v$ or $\alpha_{\rm v, total}$.  This explains why many
of the moderate-$\nhmin$ objects with low apparent virial parameter in
\autoref{fig:av}b,c,f,g mostly consist of unbound gas.
Due to all these effects, both HBR bound and unbound objects can
appear bound or unbound according to $\alpha_{v}$ and total
$\alpha_{v}$.

All of the above effects will be an issue for real clouds as well as
the structures in our simulations.  Thus, we caution that simple
estimates of gravitational energy relative to kinetic energy are
generally inadequate for assessing whether observed GMCs are genuinely
bound structures.

To determine whether observed GMCs are genuinely
bound, a similar procedure to what we have applied in this paper would
be required. That is, the first step would be to compute the
gravitational potential from all relevant material.  While three-dimensional
structure is not in  general known, previous tests have shown  that
projected surface density combined with an estimated line-of-sight depth 
is sufficient when clouds mutually  lie in a planar configuration
\citep{Gong2011}.   Inclusion of the gravitational potential from all
surrounding material is particularly important for GMCs that are found in
spiral arms, where the close proximity of clouds leads to significant
tidal effects.  

Our finding that the traditional virial parameter (\autoref{eq:virialdef}
with \autoref{eq:Egdef})
is at best an approximate
measure of boundedness has implications for interpretations of $\alpha_v$ in
observations that are otherwise quite puzzling.  For example,
\citet{Roman-Duval2010} found that GMCs identified from $^{13}$CO
Galactic Ring Survey observations have median $\alpha_v  \sim 0.5$, with mode
$\sim 0.3$.  Because a low level of kinetic energy would rapidly
lead to collapse,
it is difficult to understand how this situation could arise
unless GMCs are strongly magnetically supported, which empirically does
not seem to be the case \citep[e.g.][]{Crutcher2012,Thompson2019}.
Indeed, in purely hydrodynamic simulations, isolated clouds
that are initiated with $\alpha_v$ significantly below 1 go through a stage rapid of
contraction, such that $\alpha_v \approx 1$ by the time star formation
commences \citep{Raskutti2016}.  The low median
$\alpha_v$ in the \citet{Roman-Duval2010} observations could  be understood
if $|\Esubt{g}|$ has been overestimated by, for example, neglecting tidal effects.  

Observational surveys of nearby galaxies at $\sim 50-100$pc resolution find
values of the traditional $\alpha_v \sim 1.5-3$ for gas in resolved
structures \citep{Sun2018}.  This
suggests that most clouds are bound, which combined with the estimated
completeness of $> 50\%$ would suggest that most molecular material is
in bound  clouds.
However, in this case the low observed $\eff \sim 0.01$ for molecular gas
\citep{Utomo2018} would
be in significant tension with our finding that bound objects (HBRs) have
$\eff \sim 0.4$.  The driven-turbulence simulations of
\citet{2012ApJ...759L..27P} have similarly found $\eff \sim 0.2-0.5$ when
$\alpha_v \sim 1$.  A possible resolution is again that the traditional
observed $\alpha_v$ may overestimate boundedness by treating each cloud as
isolated.

\subsubsection{Star formation efficiency: variations and correlations}

Our results regarding the low value $\eff \sim 0.01$ of the efficiency
per free-fall time at ``average'' gas conditions is consistent with
previous observational work across a range of galaxies
\citep[e.g.][and citations within]{Evans2009,Krumholz2012,Evans2014,Lee2016,Ochsendorf2017,Utomo2018} as well as previous
numerical simulations \citep[][]{Kim2013}.  In addition, some
observations have indicated an increase of $\eff$ with density of
individual structures within given galaxies
\citep[e.g.][]{Krumholz2007,Vutisalchavakul2016}, consistent with the
trend we have identified.  Since star formation is only occurring in
the very densest regions, the variations of $\eff$ with density
threshold in a given environment, both in observations and in our
simulations, reflects the relative abundances of gas at different
densities, i.e. the density probability density function (pdf).
Analyses of the power-law portion of pdfs in Milky-Way molecular clouds
\citep[e.g.][]{Schneider2015a,Schneider2015b} 
imply a decrease of $M/\tff$ at higher density, which is compatible
with the increase of $\eff$ with density that we have found
(\autoref{fig:eff}a,b). 
The density pdf in turn reflects a ``nested'' dynamical evolution:
successively denser structures form in a hierarchical fashion, with
only a fraction of the gas at a given density experiencing net
compression by gravity and by thermal, turbulent, and magnetic
pressure to attain a higher density.  Our temporal analysis provides
evidence for hierarchical dynamics at work, in that mass histories at
varying density are offset by time delays that scale with the
gravitational free-fall time.  

Recent observations across varying galactic environments have suggested
that $\eff$ is not a function of absolute density, but of density
contrast relative to ambient levels 
\citep[e.g.][]{Garcia2012,Longmore2013,Usero2015,Gallagher2018,Querejeta2019}, although this
interpretation is complicated by uncertainties in environmental
variation of conversion factors for dense gas tracers
\citep{Shimajiri2017}.  While our present analysis considers only a
single galactic environment, we will be able to test the extent to
which $\eff$ depends on relative vs. absolute density via analysis of
additional TIGRESS simulations which have been completed for
inner-galaxy and galactic-center environments.

In addition to systematically larger $\eff$ at higher density, our
analysis shows systematically better correlations of the temporal
histories of $\SFR$ and (time-offset) histories of $\eff M/\tff$ at
higher density (\autoref{fig:time_density} and
\autoref{fig:time_others}).  This can be quantified by the systematic
decrease in $\sigma_{\Delta\SFR/\langle\SFR\rangle}$ for higher
density gas as shown in \autoref{fig:eff}.  A simulation
provides the benefit of being able to correct for the time delay between
the formation of a given defined structure and the resulting star
formation.  Since the $\SFR$ is highly variable, this time delay
produces deviations between the simultaneous $\eff M/\tff$ and $\SFR$
on the order of $t_{\mathrm{delay}}{d(\SFR)/dt}$.
For lower density gas in which
$t_{\rm delay} \sim \tff$ is long, time delays inherently
make SFRs in observations appear less correlated with the ``simultaneous''
gas mass than they really should be (as in
\autoref{fig:mbtsfr}).  The combination of the stronger inherent
correlation in amplitude variations and smaller time delays implies
that there should be less scatter in the observed statistical
correlations between $\SFR$ and mass of high density tracers in
comparison to low density tracers (assuming that the measurement of
the $\SFR$ is based on a tracer with a short timescale that does not
itself wash out the signal).  

With a sufficiently large sample of environments such that galactic
conditions can be controlled (e.g. specifying limited ranges of both
total gas and stellar surface density), and such that all phases of the star
formation cycle are well sampled for given conditions, increasingly
quantitative measures of the relationship between gas and star
formation become possible.  For example, full sampling over temporal
history can minimize effects of time delays when evaluating the
overall $\eff$ for low-density gas.  In addition, it will be possible
to quantify increases in the correlation of SFR and $M/\tff$ with
density (we measure this by a reduction in
$\sigma_{\Delta\SFR/\langle\SFR\rangle}$) while controlling for environment;
steps towards this have
already been taken \citep[e.g.][]{Gallagher2018,Jimenez2019}.
Given sufficiently
high resolution observations, it may also be possible to use analysis
of spatial correlations between high density tracers and star
formation \citep[e.g.~as in][]{Kruijssen2019} as a proxy  to measure 
temporal correlations between $\SFR$ and dense gas mass that we have
identified using simulations, thereby characterizing the bursty nature of
$\SFR$.

Finally, we remark on the relation between our work and other
theoretical/computational studies that address the relationship between
gas and star formation.  Many studies have focused exclusively
on the cold and dense ISM, because this is the material most proximate
to star formation.  With a narrower focus it is also possible to define
an idealized system with a reduced number of parameters; a minimal set
of parameters to describe gas in molecular clouds would include the
turbulent Mach number,  the ratio of the mean Alfv\'en speed to
the sound speed, and the ratio  of the Jeans length to cloud size (or
equivalently free-fall time to turbulent crossing time) \citep{Ostriker1999}.
Based on a set of idealized simulations of
this kind, with turbulence driven to maintain a fixed level, 
\citep{2012ApJ...759L..27P} proposed that $\eff$ exponentially declines
with increasing virial parameter.  As noted above, 
for moderate density threshold ($\nhmin = 10$) our fitted
coefficients are consistent with their results.  However, this is not the
case when we consider gas at higher density thresholds.  This may be because
of limited resolution at higher density thresholds in our simulations, or
because physical feedback in our simulations differs from idealized
turbulent driving, which (together with the multiphase nature) means that
all scales are not equivalent.

A class of simple theoretical models for star formation rates in
turbulent systems is predicated on the notion that there is a critical
density $\rho_{\rm crit}$, with structures at density above $\rho_{\rm
  crit}$ collapsing before they can be torn apart by ambient
turbulence \citep[e.g.][]{Krumholz2005,Padoan2011,Hennebelle2011,Federrath2012}.
These theoretical models are intended to represent idealized GMC conditions,
with gas effectively isothermal and turbulence highly supersonic; they are
therefore not immediately applicable to the present multiphase ISM simulations.
Still, it is interesting to note that our analysis does not provide evidence
that there  is a ``point of no return'' at any particular density.  Rather,
there is an order of magnitude variation in the density of bound clouds
(\autoref{fig:hist_x}g), with the probability of gas being bound and $\eff$
  both increasing with density (\autoref{fig:pbound}, \autoref{fig:error_hist}).
The present analysis does not provide information about individual cloud
lifetimes, however. Both for large-scale multiphase ISM simulations and for
smaller-scale simulations of star-forming clouds, 
numerical measurements of the lifetimes of individual structures are needed
in order to test theoretical concepts of gravoturbulent fragmentation,
and to assess whether simulations agree with observational constraints \citep[e.g.][]{Murray2011,2016ApJ...833..229L,2019MNRAS.488.1501G}. 
While some estimates of object lifetimes can be obtained via
frame-to-frame differences in structural decompositions, the most direct
way to follow evolution is via Lagrangian tracer particles.  Tracers are
commonly implemented to follow baryon cycles of gravitational
collapse and dispersal by feedback in cosmological simulations of
galaxy formation \citep[e.g.][]{Genel2013,Cadiou2019}, and for the same
reasons would be a valuable tool for future numerical studies
of  the star-forming interstellar medium.

\acknowledgments

The work was supported by the National Science Foundation under grant
AST-1713949 and NASA under grant NNX17AG26G to ECO, and grant
DGE-1148900 providing a Graduate Research Fellowship to SAM.

\appendix
\section{Structure-finding algorithm}\label{sec:algorithm}

As in the GRID-core finding algorithm \citep{Gong2011,Gong2013}, each
local minimum of the gravitational potential is associated with a
structure. The structure is composed of the material within the largest
closed isosurface containing it with a single local minimum.  All such
structures at the bottom of the hierarchy are unique.  Material within a
structure, if devoid of positive energy contributions, would collapse towards
the local potential minimum.  For some material in the structure closest to the
bounding equipotential, the thermal, kinetic, and magnetic energy might
be large enough that it cannot be considered bound to the potential minimum.

Given a potential field $\Phi$, the GRID algorithm first identifies
local minima. From each local minimum, the algorithm marches upward by
step size $\Delta\Phi$ until the contiguous region contains more than
one minimum.  The largest contour value containing only one minimum
determines the cells belonging to the structure associated with that local
minimum.

Two limitations of the algorithm are that its speed and accuracy
depend on the resolution $\Delta\Phi$. A smaller $\Delta\Phi$ ensures
that fewer cells are prematurely cut off from the structure being built,
but also
increases the number of repeated calculations of contiguous regions of
cells.
We address both limitations with an algorithm which computes structure 
membership cell by cell. It is then guaranteed that each cell is only
compared with its neighbors, and so the algorithm depends on the
number of neighbors.
This algorithm also computes the full contour tree, which can be processed
afterwards in various ways, resulting in merged objects as in our HBRs (\autoref{sec:mhbr}) or un-merged objects as in GRID.

\subsection{Algorithm Procedure}

\begin{enumerate}
\item Every cell in the 3D data set is identified with a unique positive integer ``identity''
\item The identities corresponding to each cell's neighbors are computed and optionally stored. The integer assignment is chosen so that this computation is simple.
\item The list of cell integers is sorted according to increasing $\Phi$.
\item A list of ``labels'' corresponding to the integers is initialized so that all cells are labeled as unprocessed (-1 can be used)
\item Local minima cells are labeled by their unique integer (identity), hence becoming members of their own structures, and cells in a given structure are labeled by the seed critical point of that structure.
\item Iterating over the list in order of increasing $\Phi$ (step 3), cells are labeled and assigned to structures according to rules (below) dependent only on the labels of their neighboring cells, which are easily accessed due to step 2.
\end{enumerate}

A structure is a closed isosurface containing a contiguous set of
cells with lesser $\Phi$, so the structure membership of a given cell
only depends on ``lesser neighbor'' cells with lesser $\Phi$.
Any lesser neighbor is already labeled due to steps 3 and 6,
so at any given time, the labels of the neighboring cells contain
all the information necessary to determine structure
membership. Let the label set of a cell be the unique set of labels of its
neighboring cells, ignoring the unprocessed label. 

A cell whose only lesser neighbors are members of only one structure
(label set contains exactly 1 label) 
is also a member of that one structure, and is labeled accordingly. This
is how membership propagates.

A cell with no lesser neighbors (label set is empty) must be a local minimum,
and labeled as such, as its
neighbors are all greater. If the cell is not accepted to be a
structure for any reason (e.g. boundary condition or special use
case), it can instead be assigned to a user-defined label, which will propagate as above.

A cell whose lesser neighbors are members of multiple structures (label set contains multiple elements) would define an isocontour containing all enclosed structures. This cell is a new critical point where multiple structures merge.
Hence, a new structure is defined starting from this cell.
All cells enclosed by the new structure should be relabeled to this cell's
identity.

In practice, it is
more efficient to keep track of the merger tree of the critical points, not changing previously processed cell labels. The ``local label'' for a cell corresponds to the nearest (in the tree)
lesser critical point, some of which have merged to critical points at larger $\Phi$. The label used for computing label sets is found by looking up the largest critical point in the merger tree corresponding to the ``local label.'' The combination of cell local labels and merger tree contains the necessary information to quickly access all cells belonging to any structure in the hierarchy, or to access all structures that a cell belongs to. This is how structures merge.

In this last case, any structure connected to the cell is complete,
since all cells connected to the structure with $\Phi < \Phi_{i}$ were
previously processed and added to the structure.
No other cells can be added to the structure without defining a greater
isocontour containing multiple structures, which would exactly be the new
structure defined from the critical point.
This shows that our structures
are complete and contain all viable cells, in a way that is agnostic
of choice of $\Delta\Phi$.

The computation ends when all cells are explored or when all
structures are deactivated (for example, due to a boundary condition, or if
merging structures is not allowed).
It is possible for all structures to be deactivated before all cells are explored,
which further increases the efficiency of the algorithm, because many cells
can be left un-computed. A check to
ensure active structures continue to exist can follow every structure
deactivation to minimize the number of checks.

\subsection{Strengths}

The algorithm is efficient. For $n$ cells, the algorithm requires
$\mathcal{O}(n\log{}n)$ operations to sort. For $k$ neighbors,
$\mathcal{O}(kn)$ operations are needed to compute the
neighbors. Strictly fewer than $n$ operations are required to assign a
label to each cell, because the algorithm terminates when no active
structures remain. A small amount of memory is used to keep track of
the critical point merger tree:
at most $\mathcal{O}n$.
For memory, there can be
at most $n$ labels, $kn$ neighbors (each neighbor has a 1-d index),
and $n$ values of $\Phi$. Since each cell is only accessed 1 time
during iteration, it is difficult to imagine a drastically different
scaling for the operation. A Python implementation of this algorithm can process roughly 8 million cells in a minute on a modern CPU ($256^3$ box in 2 minutes). 

In its current form, the user chooses no parameters. The algorithm
works as a black box, converting a 2-D or 3-D field into a list of
structures, their members, and their merger tree. 

The algorithm can be generally used with various cell geometries, as
long as each cell knows its neighbors.

\subsection{Extensions}

The algorithm is also relatively easy to understand, requiring very
little background, and hence easy to extend and adopt. This is because
it only aims to do a very simple task. We describe a few relevant
extensions.

The simplest extension is to apply the algorithm to the negative of a
field, to locate isocontours around maxima. This could be useful for intensity maps or density fields. 

To analyze grids with adaptive mesh refinement, computing the
neighbors of each cell is required to use the algorithm, but otherwise
can be directly used without subsampling or interpolation.

Another example is a box with sheared-periodic boundary conditions,
where the neighbors of boundary cells must be computed based upon the
shear of the box.

This should also be applicable to unstructured moving meshes.
The algorithm only needs to know which data points are neighbors.

A minimum structure size can be defined, and when two active
structures meet, an active structure which is too small is subsumed by
the larger structure. This is useful if the data has high-frequency
noise.

\bibliographystyle{aasjournal}
\bibliography{refs}{}

\end{document}